\documentclass[fleqn,10pt]{wlscirep}
\usepackage[utf8]{inputenc}
\usepackage[T1]{fontenc}

\usepackage{amsmath,amssymb,bm}
\usepackage{multirow}
\usepackage{booktabs}
\usepackage{makecell}
\usepackage{tabularx}
\usepackage{float}
\usepackage{graphicx}
\usepackage{subcaption}
\usepackage{setspace}
\newcommand{\xb}[1]{%
  \pdfliteral direct{2 Tr 0.35 w}%
  \boldsymbol{#1}%
  \pdfliteral direct{0 Tr 0 w}%
}

\title{Trusting the Inverse: Reliability-Aware Mapping for Simulation-Based Microstructure Estimation in Diffusion MRI}

\author[1,2,*]{Juan Luis Villarreal Haro}
\author[2,3]{Ileana Jelescu}
\author[1,2,3]{Jean-Philippe Thiran}
\author[1]{Jonathan Rafael-Patino}

\affil[1]{Signal Processing Laboratory (LTS5), \'Ecole Polytechnique F\'ed\'erale de Lausanne (EPFL), Lausanne, Switzerland}
\affil[2]{Radiology Department, Centre Hospitalier Universitaire Vaudois and University of Lausanne, Lausanne, Switzerland}
\affil[3]{CIBM Center for Biomedical Imaging, Lausanne, Switzerland}

\affil[*]{juan.villarrealharo@epfl.ch}


\begin{abstract}

Diffusion-weighted MRI can probe tissue microstructure non-invasively, however interpreting microstructural parameter estimates remains challenging due to the intrinsic ambiguity of the inverse problem. While simulation-based approaches can incorporate increasingly realistic tissue models, they still lack voxel-wise characterisation of the reliability and degeneracy of the inferred parameters. Here, we introduce a reliability framework for simulation-based microstructure estimation based on three complementary scores that identify distinct sources of unreliability in the estimation process: out-of-distribution signals, local signal mismatch, and parameter degeneracy. The framework was implemented using a Monte Carlo dictionary of 1{,}050 synthetic voxels generated from geometrically realistic substrates, with parameter ranges grounded in electron microscopy measurements of rat corpus callosum. The dictionary spans biologically plausible axon radii (0.25-0.85~$\mu$m), microscopic angular spread (0-10$^\circ$), packing densities (60-92\%), and intrinsic diffusivities (1.75-3.0~$\mu$m$^2$/ms). On synthetic data, the resulting Reliability Index correlated with actual estimation error (Spearman $\rho = -0.742$) and distinguished between extrapolation, poor local interpolation, and parameter degeneracy. Applied to in vivo corpus callosum DW-MRI in rat (256~voxels, four animals) and human (MGH-USC HCP, 18{,}765~voxels, nine subjects), 91\% and 73\% of voxels respectively exceeded $R > 0.5$. These results demonstrate how reliability-aware analysis can support the interpretation and future development of simulation-based diffusion MRI microstructure models.

\end{abstract}

\begin{document}
\onehalfspacing
\flushbottom
\maketitle
\thispagestyle{empty}

% --- Main sections ---
% introduction.tex — ~1,000 words, 5 paragraphs, \section*{Introduction}

\section*{Introduction}

Diffusion-weighted magnetic resonance imaging (DW-MRI) provides non-invasive sensitivity to tissue microstructure through the displacement of water molecules in biological tissue\cite{LeBihan2024,Basser1994}. In brain white matter, biophysical models use these diffusion signals to estimate microstructural properties such as axon diameter, neurite density, and orientation dispersion\cite{Zhang2012,Alexander2010,Assaf2005,Assaf2008}. However, white matter organisation remains considerably more complex than the assumptions underlying most signal models, with axonal bundles exhibiting broad diameter distributions, non-cylindrical geometries, and substantial morphological variability along their trajectories\cite{Andersson2021,lee2019along}. Despite these limitations, DW-MRI-derived microstructural parameters are increasingly investigated as candidate biomarkers for neurodegeneration, demyelination, and abnormal development\cite{LeBihan2024}. Interpreting these estimates therefore requires not only parameter recovery, but also an assessment of how reliably the underlying tissue configuration is constrained by the measured signal.

To better capture the structural complexity of white matter, diffusion MRI increasingly relies on simulation-based approaches that move beyond simplified analytical representations of tissue microstructure\cite{Neuman1974,VanGelderen1994,Soderman1995}. Monte Carlo diffusion simulations model water displacement directly within three-dimensional substrates, enabling the generation of synthetic signals from geometries that incorporate irregular axonal cross-sections, heterogeneous packing, and realistic spatial organisation. These simulations provide a flexible framework for exploring microstructural models and constructing signal dictionaries for simulation-based parameter estimation.

Simulation-based estimation frameworks commonly rely on dictionaries of synthetic diffusion signals generated across a range of microstructural configurations\cite{Rensonnet2019,Daducci2015}. In this context, the realism and biological grounding of the simulated substrates become critical for ensuring meaningful overlap between synthetic and experimental signal spaces. Recent substrate generators have enabled increasingly realistic representations of white matter organisation, including heterogeneous packing, non-cylindrical axonal geometries, and histology-informed diameter distributions\cite{Villarreal-Haro2023}. Among these, the CACTUS framework enables large-scale dictionary construction in computational domains of $(120~\mu\text{m})^3$ with packing densities exceeding 90\%, making it suitable for simulation-based microstructure estimation under realistic geometric constraints.

However, increasing model realism does not eliminate the intrinsic ambiguity of the inverse problem. Distinct tissue configurations may still produce nearly identical diffusion signals, leading to parameter degeneracy and multiple plausible solutions\cite{Jelescu2016,Novikov2018}. As simulation-based models become increasingly detailed, interpreting the reliability of the inferred parameters becomes equally important. Existing uncertainty estimation approaches, including Bayesian inference\cite{Cranmer2020,Jallais2024} and bootstrap resampling\cite{Chung2006}, can quantify uncertainty in different forms, but they generally do not identify the specific source of unreliability or indicate which aspects of the acquisition, dictionary, or model representation limit the estimation process. In particular, existing approaches do not distinguish whether unreliable estimates arise from out-of-distribution signals, insufficient local signal representation, or intrinsic parameter degeneracy.

In this work, we introduce a reliability framework for simulation-based microstructure estimation built around these three sources of unreliability. We construct a dictionary of 1{,}050 Monte Carlo voxels using CACTUS substrates coupled with MC-DC diffusion simulations\cite{Rafael-Patino2020,Rafael-Patino2018}, with parameter ranges grounded in electron microscopy measurements of rat corpus callosum\cite{pesaresi2015axon} and acquisition parameters matched to the in vivo protocol. The framework is then applied to in vivo DW-MRI of rat corpus callosum acquired using the time-dependent diffusion MRI protocol\cite{Jelescu2022}, and to human corpus callosum from the MGH-USC Human Connectome Project dataset\cite{Setsompop2013,Fan2016}. To characterise estimation reliability at the voxel level, we introduce the Reliability Index, composed of three complementary scores: the Outlier-Detection score ($S_{\text{out}}$), the Signal-Matching score ($S_{\text{match}}$), and the Parameter-Degeneracy score ($S_{\text{deg}}$). Together, these scores provide an interpretable characterisation of estimation reliability relative to the simulated dictionary and acquisition protocol.

% methods.tex — unlimited words, with subheadings

\section*{Methods}

% ----------------------------------------------------------------------------
\subsection*{In vivo rat corpus callosum DW-MRI}
% ----------------------------------------------------------------------------

%Our framework requires experimental data from a tissue whose microstructure has been widely studied in histological measurements. 

To ground the simulation parameter ranges in biological measurements, we used a tissue whose microstructure has been extensively characterised with electron microscopy. We retrospectively used rat diffusion MRI datasets acquired in the context of NEXI\cite{Jelescu2022}. The dataset comprises four in vivo whole-brain DW-MRI acquisitions from Wistar rats. All animals were scanned on a 9.4~T preclinical system using a PGSE sequence. Rats~1-3 were acquired at $0.2 \times 0.2$~mm$^2$ in-plane resolution with 0.5~mm slices, and Rat~4 at $0.25 \times 0.25$~mm$^2$ with 0.8~mm slices. Each acquisition sampled 7-8 non-zero $b$-value shells spanning 1{,}000-10{,}000~s/mm$^2$ with 24 gradient directions per shell and $\delta = 4.5$~ms, at diffusion times $\Delta = 12, 20, 30, 40$~ms for Rats~1-3 and $\Delta = 11, 25, 45$~ms for Rat~4. For each animal we used the longest available diffusion time ($\Delta = 40$~ms for Rats~1-3, $\Delta = 45$~ms for Rat~4) to maximise sensitivity to microstructural restriction. We chose to focus on the rat corpus callosum, a well-characterised white matter structure commonly used for microstructural validation studies.

%(PGSE) sequence on a 9.4\,T preclinical scanner. Rats~1--3 used an in-plane resolution of $0.2 \times 0.2$~mm$^2$ and 0.5~mm slice thickness; Rat~4 used $0.25 \times 0.25$~mm$^2$ and 0.8~mm. The acquisition used multi-shell, multi-diffusion-time sampling with 7--8 non-zero $b$-value shells (1,000--10,000~s/mm$^2$) and 24 gradient directions per shell. Rats~1--3 used $b$-values of [1000, 2500, 4000, 5500, 7000, 8500, 10\,000]~s/mm$^2$ at diffusion times $\Delta = [12, 20, 30, 40]$~ms, while Rat~4 used [1000, 2500, 5000, 6000, 7000, 8000, 9000, 10\,000]~s/mm$^2$ at $\Delta = [11, 25, 45]$~ms. For all analyses, we used the highest available diffusion time per animal ($\Delta = 40$~ms for Rats~1--3, $\Delta = 45$~ms for Rat~4) to maximise sensitivity to microstructural restrictions. 

%To select well-aligned white matter voxels and reduce partial volume effects\cite{kumazawa2010partial}, we applied an FA threshold of 0.70 to the manually segmented corpus callosum mask, resulting in 65~voxels across all four animals (Rat\,1:~15, Rat\,2:~26, Rat\,3:~18, Rat\,4:~6). Given this sample size, the present study constitutes a proof-of-concept demonstration rather than a statistically powered population analysis.
All voxels of the manually segmented corpus callosum mask were analysed, giving 256~voxels across the four animals (Rat\,1:~66, Rat\,2:~89, Rat\,3:~63, Rat\,4:~38). No fractional-anisotropy threshold was applied, so voxels near the mask boundary remain subject to partial volume effects\cite{kumazawa2010partial}. Given this sample size, the present study constitutes a proof-of-concept demonstration rather than a statistically powered population analysis.

% end removed acquisition table

% ----------------------------------------------------------------------------
\subsection*{In vivo human corpus callosum DW-MRI}
% ----------------------------------------------------------------------------

To evaluate the framework on a larger in vivo dataset, we used publicly available human brain DW-MRI from the MGH-USC Human Connectome Project Adult Diffusion data\cite{Setsompop2013,Fan2016}. From the 35 subjects in this release we analysed nine, acquired on a 3\,T Connectom scanner with a maximum gradient strength of 300~mT/m at 1.5~mm isotropic resolution. The protocol used $\Delta = 21.8$~ms, $\delta = 12.9$~ms, and $b$-values of 1{,}000, 3{,}000, 5{,}000, and 10{,}000~s/mm$^2$. Spherical-mean signals were computed at these $b$-values for matching against the simulated dictionary. Corpus callosum masks were obtained from FreeSurfer segmentations registered to the diffusion space, yielding 18{,}765 voxels across the nine subjects.

% --- Table: Acquisition parameters (Display item 8) ---
% \begin{table}[ht]
% \centering
% \small
% \renewcommand{\arraystretch}{1.4}
% \begin{tabular}{@{}llccl@{}}
% \toprule
% \textbf{Category} & \textbf{Parameter} & \textbf{Range / Value}
%   & \textbf{$N$} & \textbf{Notes} \\
% \midrule
% \multirow{5}{*}{\textbf{Real Voxels}}
%   & Volumes              & --                        & 4          & Corpus callosum \\
%   & Voxels per mask      & --                        & $\sim$66   & Before thresholding \\
%   & Voxels with FA $>$ 0.7 & 6--26 per mask          & 65         & Total \\
%   & In-plane resolution (mm$^2$) & $(0.2)^2$, $(0.25)^2$ & -- & \\
%   & Slice thickness (mm) & 0.5, 0.8                  & --         & \\
% \midrule
% \multirow{6}{*}{\textbf{Diffusion Protocol}}
%   & $b$-values (s/mm$^2$)  & 1000--10\,000   & 7--8 & Variable spacing \\
%   & Gradient directions    & --              & 24   & per $b$-value/shell \\
%   & $\Delta$ (ms)          & 11--45          & 3--4 & Per volume \\
%   & $\delta$ (ms)          & 4.5             & --   & \\
%   & TE (ms)                & 50, 52, 58      & --   & \\
%   & TR (ms)                & 2500--3000      & --   & Volume-dependent \\
% \bottomrule
% \end{tabular}
% \caption{Acquisition parameters for the NEXI in~vivo dataset from Wistar rat corpus callosum. Total analysed: 65~voxels across 4~volumes after FA thresholding. Acquisition protocols vary slightly between volumes.}
% \label{tab:acquisition}
% \end{table}
% ----------------------------------------------------------------------------
\subsection*{Histology-informed substrate generation with CACTUS}
% ----------------------------------------------------------------------------

We generated geometrically realistic three-dimensional axonal substrates using the CACTUS framework\cite{Villarreal-Haro2023}, with parameter ranges grounded in histological measurements of rat corpus callosum. We parameterised axon radius distributions by their mean radius, sampled from 0.25 to 0.85~$\mu$m in 0.1~$\mu$m increments (7~values), encompassing the range reported in histological studies of the rat corpus callosum\cite{pesaresi2015axon} and, in the human corpus callosum, the outer radii of myelinated axons (approximately 0.4 to 0.7~$\mu$m at a $g$-ratio of 0.7)\cite{Aboitiz1992,Caminiti2013}. For all radii distributions, we fixed the mean g-ratio at 0.7, consistent with established corpus callosum measurements\cite{Stikov2015}. The microscopic angular spread ($\mu\theta$), which is a substrate parameter defined as the angular spread of individual axon trajectories about the mean bundle axis within a voxel, was varied from 0$^\circ$ (parallel axons) to 10$^\circ$ (tortuous intertwined trajectories) in 2.5$^\circ$ steps (5~values), consistent with the low microscopic angular spread of corpus callosum fibres\cite{lee2019along,Mollink2017}. Because the spherical-mean representation used for matching (Eq.~\ref{eq:spm}) removes the fibre orientation distribution, varying $\mu\theta$ across the dictionary tests the robustness of the remaining parameter estimates (axon radius, ICVF, and intrinsic diffusivity) to intra-voxel orientation dispersion, rather than recovering orientation dispersion from the powder-averaged signal. We similarly varied packing density (ICVF) by sampling from 60\% to 92\% in 8\% increments (5 values), spanning the intracellular volume fraction of living brain tissue (60-95\%) measured by super-resolution imaging\cite{Tonnesen2018}. To minimise sampling bias, each substrate occupied a cubic computational domain of $(120~\mu\text{m})^3$\cite{Rafael-Patino2020}.
%The spherical mean removes the fibre orientation distribution and isolates the microenvironment kernel\cite{Kaden2016}; $\mu\theta$ is a property of that kernel, since axons deviating from the bundle axis in a densely packed substrate also deviate individually from a straight trajectory.

The 175~physical substrates correspond to $7 \times 5 \times 5 = 175$ unique combinations of axon radius, microscopic angular spread, and packing density. Combined with 6 intrinsic diffusivity values (see below), this results in $175 \times 6 = 1{,}050$ unique voxels.

% ----------------------------------------------------------------------------
\subsection*{Diffusion MRI signals via Monte Carlo simulations}
% ----------------------------------------------------------------------------

%With the three-dimensional substrate meshes in place, the next step is generating their corresponding DW-MRI signals. 

Diffusion-weighted signals were simulated using the MC-DC framework\cite{Rafael-Patino2020,Rafael-Patino2018} within the generated CACTUS substrates. Particles were initialised exclusively within the intra- and extra-axonal compartments, excluding the myelin space.

The simulated signal was computed as the volume-weighted sum of intra- and extra-axonal contributions:

\begin{equation}
  S = S_{\mathrm{intra}} \, \mathrm{AVF}
    + S_{\mathrm{extra}} \, \mathrm{ECVF},
  \label{eq:signal}
\end{equation}
where $S_{\mathrm{intra}}$ and $S_{\mathrm{extra}}$ denote normalised intra- and extra-axonal signal attenuations. The axonal (AVF) and myelin (MVF) volume fractions sum to the intra-cellular volume fraction, $\mathrm{ICVF} = \mathrm{AVF} + \mathrm{MVF}$\cite{Stikov2015}, and the extra-cellular volume fraction is $\mathrm{ECVF} = 1 - \mathrm{ICVF}$; myelin signal is excluded from the simulation because of the short $T_2$ of myelin water\cite{MacKay1994,AlonsoOrtiz2015}, so only the axonal and extra-axonal compartments enter the equation. Our simulations do not impose compartment-specific $T_2$ relaxation, so any $e^{-\mathrm{TE}/T_2}$ factor is common to both compartments and cancels in the $b_0$ normalisation. The compartment weight in our model is therefore the geometric axonal volume fraction (ICVF), rather than the $T_2$-weighted relative signal fraction of the Standard Model and NEXI\cite{Novikov2018,Jelescu2022}.

We modelled water diffusion in both compartments with a single free-diffusion coefficient $D$, sampled at six values (1.75, 2.0, 2.25, 2.5, 2.75, and 3.0~$\mu$m$^2$/ms) to cover the physiological range of intrinsic diffusivity in neural tissue\cite{mills1973self,dhital2019intraaxonal}. The diffusion signals were simulated by replicating each dataset's in vivo acquisition protocol with a PGSE sequence. For the rat data, we reproduced each animal's acquisition: 24 gradient directions per shell at its multi-shell $b$-values (up to 10{,}000~s/mm$^2$), its diffusion time, and $\delta = 4.5$~ms. For the human data, we reproduced the MGH-USC HCP protocol: $b$-values of 1{,}000, 3{,}000, 5{,}000, and 10{,}000~s/mm$^2$ at its single diffusion time. Measured and simulated spherical means were therefore compared on identical protocols in each dataset. Simulations used a particle density of 1~particle/$\mu$m$^3$ and 1{,}500 time steps\cite{Hall2009}. To avoid periodic-domain boundary artefacts, we extracted signals from a centred $(100~\mu\text{m})^3$ subvolume, discarding the 10~$\mu$m border on each side.

% ----------------------------------------------------------------------------
\subsection*{Parameter estimation}
% ----------------------------------------------------------------------------

% To match each in vivo voxel to its closest dictionary entry, we needed a per-voxel signal representation: simple enough to compute reliably, yet informative enough to distinguish microstructural configurations.  The spherical mean is our first candidate. Averaging over gradient directions removes orientation dependence and leaves a rotationally invariant summary that depends only on tissue microstructure. For each $b$-value shell, we computed:

Voxel matching was performed using the spherical mean representation of the diffusion signal. By averaging across gradient directions, the spherical mean provides a rotationally invariant signal representation that enables direct comparison between measured and simulated voxels independently of fibre orientation. For each $b$-value shell, we computed:

\begin{equation}
  \bar{S}(b)
    = \frac{1}{N_{\text{dir}}}
      \sum_{i=1}^{N_{\text{dir}}} S(b,\, \mathbf{g}_i).
  \label{eq:spm}
\end{equation}

We quantified voxel similarity using the logarithmic mean absolute error (log-MAE):
\begin{equation}
  d_{\text{log-MAE}}
    = \frac{1}{N_b}
      \sum_{i=1}^{N_b}
        \bigl|
          \log\!\bigl(\bar{S}_{\text{real}}(b_i) + \eta\bigr)
          - \log\!\bigl(\bar{S}_{\text{synth}}(b_i) + \eta\bigr)
        \bigr|,
  \label{eq:logmae}
\end{equation}
where $\eta = 10^{-6}$ prevents numerical instability at highly attenuated signals. The logarithmic transform increases sensitivity to relative differences at high $b$-values, where signal attenuation is strongest.

% ----------------------------------------------------------------------------

% ----------------------------------------------------------------------------
%Given a dictionary of signal-parameter pairs, the simplest estimator is to find the entries most similar to the test signal and read off their parameters. We formalised this with exponentially weighted $K$-nearest neighbours (KNN) regression, which avoids fitting an explicit forward model and instead interpolates directly from the dictionary. We chose $K = 10$ because it is large enough for a stable weighted average while staying close to the test signal; smaller values amplify noise, while larger values pull in distant, less-relevant entries. The sharpness parameter $\alpha = 10$ concentrates most weight on the two or three closest neighbours, with a small contribution from the rest; pilot experiments showed no meaningful change in accuracy for $\alpha \in [5, 20]$. For each test voxel, we identified the $K$ nearest dictionary entries by log-MAE distance, yielding a neighbour set $\mathcal{N} = \{(\bar{S}_i, \boldsymbol{\theta}_i, d_i)\}_{i=1}^{K}$, where $\boldsymbol{\theta}_i = (r_i, \kappa_i, f_i, D_i)$ is the known parameter vector (axon radius, orientation dispersion, ICVF, intrinsic diffusivity). We then weighted them by an exponential kernel:

Parameter estimation was performed using exponentially weighted $K$-nearest neighbours (KNN) regression on the dictionary signal representations. For each test voxel, we identified the $K = 10$ nearest dictionary entries according to the log-MAE distance, resulting in a neighbour set $\mathcal{N} = \{(\bar{S}_i, \boldsymbol{\theta}_i, d_i)\}_{i=1}^{K}$, where $\boldsymbol{\theta}_i = (r_i, \mu\theta_i, f_i, D_i)$ denotes the parameter vector (axon radius, microscopic angular spread, ICVF, intrinsic diffusivity) associated with dictionary entry $i$. We used $K = 10$ neighbours and a kernel sharpness parameter $\alpha = 10$, which provided stable estimates across the explored parameter range. Neighbour contributions were weighted using an exponential kernel:

\begin{equation}
  w_i
    = \frac{
        \exp\!\bigl(-\alpha \cdot (d_i - d_{\min})\bigr)
      }{
        \sum_{j=1}^{K} \exp\!\bigl(-\alpha \cdot (d_j - d_{\min})\bigr)
      },
  \label{eq:weights}
\end{equation}
where $d_{\min} = \min_i d_i$. The point estimate is the weighted mean $\hat{\boldsymbol{\theta}} = \sum_{i=1}^{K} w_i \boldsymbol{\theta}_i$.

Marginal 95\% confidence intervals were estimated via bootstrap resampling of the neighbour set (500 iterations). The bias-corrected weighted covariance matrix characterises the full joint uncertainty structure:
\begin{equation}
  \Sigma
    = \frac{1}{1 - V_2}
      \sum_{k=1}^{K} w_k
        \bigl(\boldsymbol{\theta}_k - \hat{\boldsymbol{\theta}}\bigr)
        \bigl(\boldsymbol{\theta}_k - \hat{\boldsymbol{\theta}}\bigr)^\top,
  \qquad
  V_2 = \sum_{k=1}^{K} w_k^2,
  \label{eq:covariance}
\end{equation}
with parameters normalised by their range $\Delta\theta^{(j)}$ to yield the normalised covariance $\Sigma^*_{ij} = \Sigma_{ij} / (\Delta\theta^{(i)} \Delta\theta^{(j)})$.

% ----------------------------------------------------------------------------
\subsection*{Three reliability scores: Outlier-Detection, Signal-Matching, and Parameter-Degeneracy}
% ----------------------------------------------------------------------------

%Knowing that an estimate comes from nearby dictionary entries is not enough: the neighbours could be outliers, the match could be poor, or the parameters could be degenerate. 

Neighbour proximity in signal space alone does not fully characterise estimation reliability. Similarity-based matching may be unreliable because the measured signal lies outside the dictionary support, because the local signal representation is insufficiently sampled, or because multiple microstructural configurations produce nearly indistinguishable signals. We therefore designed three targeted scores that separately quantify each source of unreliability and combine them into a single Reliability Index (Table~\ref{tab:reliability}). Each score maps a raw quantity to the interval $[0,\,1]$ via a monotone decreasing function of the form $1/\bigl(1 + (x/\beta)^{\alpha}\bigr)$, where $\beta$ is the half-score point ($S = 0.5$ at $x = \beta$) and $\alpha$ controls the steepness of the transition.

\paragraph{Outlier-Detection score ($S_{\text{out}}$).}
The aim of $S_{\text{out}}$ is to detect unwanted extrapolation effects: whether the test signal lies within the signal space covered by the dictionary or not. We use the Local Outlier Factor
(LOF)\cite{breunig2000lof}, which compares the local density around the test signal
to the local density of the surrounding dictionary entries. LOF~$\approx 1$ indicates
the test signal sits in a region as densely sampled as its neighbours (in-distribution);
values increasingly above~1 indicate progressively sparser coverage, meaning the
dictionary does not represent that part of signal space:
\begin{equation}
  S_{\text{out}}
    = \frac{1}{1 + \bigl(\lambda / \beta_1\bigr)^{\alpha_1}},
  \qquad
  \lambda = \max(\text{LOF} - 1,\; 0).
  \label{eq:S_out}
\end{equation}
The LOF was computed with $k = 10$ neighbours on the dictionary signal representations.
Low $S_{\text{out}}$ values indicate that the test signal falls outside the region of signal space represented by the dictionary, leading to unreliable extrapolation. This suggests insufficient dictionary coverage or missing substrate features.

\paragraph{Signal-Matching score ($S_{\text{match}}$).}
The aim of $S_{\text{match}}$ is to detect poor local interpolation. Even when a signal is within the dictionary's range, the KNN reconstruction $\hat{S}_{\mathcal{N}} = \sum_{i \in \mathcal{N}} w_i \bar{S}_i$ may not reproduce it if the dictionary is too sparse nearby. We define the normalised residual $\varepsilon = \|\bar{S}_{\text{test}} - \hat{S}_{\mathcal{N}}\| / \sigma_{\text{dict}}$, where $\sigma_{\text{dict}}$ is the standard deviation of all dictionary signals, and map it to:
\begin{equation}
  S_{\text{match}}
    = \frac{1}{1 + \bigl(\varepsilon / \beta_2\bigr)^{\alpha_2}}.
  \label{eq:S_match}
\end{equation}
Low $S_{\text{match}}$ values indicate that the local dictionary neighbourhood cannot accurately reconstruct the target signal, suggesting insufficient local sampling density or limitations in the signal representation.

\paragraph{Parameter-Degeneracy score ($S_{\text{deg}}$).} Finally,
$S_{\text{deg}}$ aims to detect parameter degeneracy: a signal that is well reconstructed while its neighbours span a broad parameter region may indicate a locally ill-posed inverse problem. The degeneracy arises from two sources: \emph{representation degeneracy} (the chosen signal representation discards discriminative information, e.g.\ powder-averaging the directional acquisitions into a spherical mean sets aside orientation contrast) or \emph{protocol degeneracy} (the acquisition lacks contrast to separate configurations). We quantify the spread of parameter estimates as the volume of the neighbour ellipsoid in normalised parameter space, taken as the generalised standard deviation of the covariance $\Sigma^*$ ($n$ parameters) with a resolution floor $\tau$ added to its diagonal:
\begin{equation}
  \nu = |\det(\Sigma^* + \tau I)|^{1/2n}
      = \left( \prod_{i=1}^{n} (\lambda_i + \tau) \right)^{1/2n},
  \qquad
  S_{\text{deg}}
    = \frac{1}{1 + \bigl(\nu / \beta_3\bigr)^{\alpha_3}}.
  \label{eq:S_deg}
\end{equation}
Here $\lambda_i$ are the eigenvalues of $\Sigma^*$ (the variances along its principal axes, expressed as a fraction of the parameter range) and $I$ is the identity matrix, so $\nu$ is the geometric mean of the regularised standard deviations $\sqrt{\lambda_i + \tau}$. Adding $\tau$ to the diagonal keeps the determinant bounded below by $\tau^{n}$ when the neighbour set is rank-deficient in one or more parameters, so the product of eigenvalues survives rather than collapsing to zero. We set $\tau = 0.10$, approximately the median normalised spacing of the dictionary grid.
% When $S_{\text{deg}}$ is low, the neighbour signals closely match the target but their parameters span a broad region: multiple configurations produce nearly identical signals. The covariance ellipses are elongated, revealing which parameter dimensions are degenerate. This points to a richer signal representation that recovers the discarded contrast, or an acquisition protocol with additional orthogonal sensitivity.
Low $S_{\text{deg}}$ values indicate that multiple parameter configurations remain compatible with the observed signal despite accurate local signal reconstruction. This degeneracy may arise from limitations in the signal representation or insufficient contrast in the acquisition protocol.

\paragraph{Per-parameter precision.}
While $S_{\text{deg}}$ summarises the joint neighbour spread into a single score, the diagonal of the normalised covariance resolves it per parameter. For parameter $j$, the marginal standard deviation of the retained neighbours, as a fraction of the parameter range, is $\sqrt{\Sigma^*_{jj}} = \sigma_j / \Delta\theta^{(j)}$, and we report a per-parameter precision
\begin{equation}
  P_j = \max\!\bigl(0,\; 1 - 2\sqrt{\Sigma^*_{jj}}\bigr),
  \label{eq:precision}
\end{equation}
the fraction of the parameter's plausible range lying outside its $\pm 1\sigma$ neighbour interval (interval width $2\sigma_j$). $P_j = 1$ marks a parameter pinned to a negligible fraction of its range, $P_j = 0$ one whose neighbour spread fills it. Whereas $S_{\text{deg}}$ aggregates the eigenvalues $\lambda_i$ of $\Sigma^*$ (spread along the principal axes of the joint uncertainty), $P_j$ reads its diagonal (spread along each parameter axis), attributing the overall degeneracy to individual parameters. $P_j$ is reported with the estimates in the voxel-level figures (Figures~\ref{fig:invivo_high_R}-\ref{fig:invivo_low_Sout}) and identifies which parameters a low $S_{\text{deg}}$ is driven by.

\paragraph{Reliability Index.}
We combine the three scores as their geometric mean:
\begin{equation}
  R = \bigl(
        S_{\text{out}} \times S_{\text{match}} \times S_{\text{deg}}
      \bigr)^{1/3}.
  \label{eq:combined_R}
\end{equation}
The geometric mean penalises low values in any individual score, ensuring sensitivity to all three sources of unreliability. We classify estimates into three quality tiers: reliable ($R > 0.60$), moderate ($0.40 \leq R \leq 0.60$), or unreliable ($R < 0.40$). These thresholds partition the synthetic validation results into groups with distinct estimation error distributions. We apply the same reliable-tier boundary to the individual scores: a score above 0.6 (60\%) indicates that its corresponding condition (dictionary support, signal matching, or parameter identifiability) is met. The Reliability Index characterises reliability within the chosen simulation and acquisition framework.

The sigmoid parameters $(\alpha_j, \beta_j)$ were calibrated separately for each dataset by maximising the separation between corpus callosum voxels and the surrounding complement region, calibrating on a subset of volumes and validating on the remainder (rat: Rats~3-4 for calibration, Rats~1-2 for validation; human: Subjects~6-9 for calibration, Subjects~1-5 for validation). For the rat protocol we used $\beta_1 = 4.872$, $\alpha_1 = 2$, $\beta_2 = 0.172$, and $\alpha_2 = 5$; for the human protocol, $\beta_1 = 6.0$, $\alpha_1 = 2$, $\beta_2 = 0.144$, and $\alpha_2 = 5$. The degeneracy sigmoid used $\beta_3 = 1.0$ and $\alpha_3 = 2$ in both datasets, since the generalised covariance volume is defined in range-normalised units and is therefore directly comparable across protocols.
% (removed from main text per author decision: the dictionary is described as the full 1{,}050-voxel set) Dictionary entries with an intra-cellular volume fraction below 0.6 were excluded before matching.

% --- Reliability framework summary table ---
\begin{table}[ht]
\caption{\textbf{Three scores for reliability assessment.} Each score targets a distinct source of unreliability and points to a targeted improvement.}
\label{tab:reliability}

\begin{center}
  \small
  \renewcommand{\arraystretch}{1.4}
  \setlength{\tabcolsep}{6pt}
  \begin{tabularx}{\linewidth}{@{} l
    >{\centering\arraybackslash}X
    >{\centering\arraybackslash}X
    >{\centering\arraybackslash}X @{}}
  \toprule
  & $\boldsymbol{S_{\textbf{out}}}$
  & $\boldsymbol{S_{\textbf{match}}}$
  & $\boldsymbol{S_{\textbf{deg}}}$ \\[-0.3em]
  & {\scriptsize OUTLIER DETECTION}
  & {\scriptsize SIGNAL MATCHING}
  & {\scriptsize PARAM.\ DEGENERACY} \\
  \midrule
  \textbf{Question} &
    Is the signal inside the dictionary? &
    Can the neighbours reproduce the signal? &
    Is the estimate well-constrained? \\[0.6em]
  \textbf{Score} &
    $\dfrac{1}{1 + \bigl(\xb{\lambda}\,/\,\beta_1\bigr)^{\!\alpha_1}}$ &
    $\dfrac{1}{1 + \bigl(\xb{\varepsilon}\,/\,\beta_2\bigr)^{\!\alpha_2}}$ &
    $\dfrac{1}{1 + \bigl(\xb{\nu}\,/\,\beta_3\bigr)^{\!\alpha_3}}$ \\[1.0em]
  {\small\itshape where} &
    {\small $\xb{\lambda} = \max(\text{LOF}\!-\!1,\;0)$} &
    {\small $\xb{\varepsilon} = \|\bar{s}_{\text{t}}-\hat{s}_{\mathcal{N}}\|\,/\,\sigma_{\text{d}}$} &
    {\small $\xb{\nu} = |\det(\Sigma^* + \tau I)|^{1/2n}$} \\[0.6em]
  \textbf{Detects} &
    Extrapolation: signal outside dictionary coverage &
    Interpolation: neighbours cannot reproduce the signal shape &
    Degeneracy: similar signals from distinct configurations \\[0.4em]
  \midrule
  \multicolumn{4}{c}{%
    \begin{tabular}{@{}c@{}}
      \\[-0.8em]
      $\displaystyle
        \boxed{\;\;
          R
            = \Bigl(S_{\text{out}}
              \times S_{\text{match}}
              \times S_{\text{deg}}\Bigr)^{1/3}
        \;\;}
      $ \\[0.2em]
      {\small Geometric mean; each score targets a different source of unreliability.} \\[0.3em]
    \end{tabular}%
  } \\
  \bottomrule
  \end{tabularx}
\end{center}

  \vspace{0.3em}
  \begingroup
  \footnotesize
  \noindent
  $\bar{s}_{\text{t}} = \text{test signal}$\,;\;\;
  $\hat{s}_{\mathcal{N}}=\textstyle\sum_{i\in\mathcal{N}}w_i\bar{s}_i$\,;\;\;
  $\sigma_{\text{d}} = \operatorname{std}(S_{\text{dict}})$\,;\;\;
  $\Sigma^*=\text{normalised w.\ cov.\ of }
    \{\boldsymbol{\theta}_i\}_{i\in\mathcal{N}}$
  \endgroup
\end{table}

% ----------------------------------------------------------------------------
\subsection*{Self-validation of the Reliability Index on the synthetic dictionary}
% ----------------------------------------------------------------------------

%Before applying the Reliability Index to in vivo data, we ran it on data where the true parameters are known: the synthetic dictionary itself. We used a leave-one-out scheme: each of the 1,050 voxels takes a turn as the test signal while the remaining 1,049 form the dictionary, ensuring no overlap between the test signal and the estimation set.  

The Reliability Index was self-validated using a leave-one-out analysis on the synthetic dictionary. Each of the 1{,}050 synthetic voxels was treated as a test signal while the remaining 1{,}049 entries formed the estimation dictionary, ensuring complete separation between test and reference signals. Rician noise was added to each simulated measurement before spherical-mean averaging, using the DIPY implementation\cite{Garyfallidis2014}. The noise standard deviation was set from the signal-to-noise ratio of the non-diffusion-weighted signal, $\sigma = S(b{=}0)/\mathrm{SNR}$, and the noisy magnitude signal was formed as $\sqrt{(S + n_1)^2 + n_2^2}$, where $n_1$ and $n_2$ are independent zero-mean Gaussian variables of standard deviation $\sigma$ representing the real and imaginary channels. Noise was added at SNR levels of $\infty$ (noiseless), 400, 200, 100, 70, 50, 40, 30, and 25, resulting in a total of 9{,}450 test cases. For each noisy signal, parameter estimation and reliability scoring were performed using the exponentially weighted KNN framework ($K = 10$, $\alpha = 10$) against the clean dictionary. Low-reliability cases were additionally grouped according to the dominant source of unreliability, defined by the minimum of $S_{\text{out}}$, $S_{\text{match}}$, and $S_{\text{deg}}$. The predictive value was quantified using the Spearman rank correlation between reliability scores and mean relative parameter error across all test cases, as the variables did not follow a normal distribution.

% % ----------------------------------------------------------------------------
% \subsection*{Spearman correlation and summary statistics}
% % ----------------------------------------------------------------------------

% We used Spearman rank correlation (two-tailed) to assess the monotonic relationship between reliability scores and estimation error, as neither variable follows a normal distribution. We report parameter estimates as medians with interquartile ranges (IQR), appropriate for the small sample sizes per animal. We performed all statistical analyses using Python (SciPy v1.11).

\subsection*{Complement-mask analysis}

To assess the behaviour of the reliability framework inside and outside the simulated tissue domain, we performed the estimation procedure on voxels inside and outside the corpus callosum mask. The complement region was defined as all voxels within a bounding box expanded by two voxels around the corpus callosum mask, excluding the mask itself. This region contains tissue types and partial-volume configurations not explicitly represented in the dictionary.

% results.tex — ~1,500 words, with subheadings

\section*{Results}

We first evaluated the behaviour of the proposed reliability framework under controlled synthetic conditions, where ground-truth microstructural parameters are available. This analysis assessed how the three reliability scores behave across noise levels and how well the combined Reliability Index reflects actual estimation error.

% ----------------------------------------------------------------------------
\subsection*{Self-validation of the Reliability Index across SNR levels}
% ----------------------------------------------------------------------------

\begin{figure}[H]
  \centering
  \includegraphics[width=\linewidth]{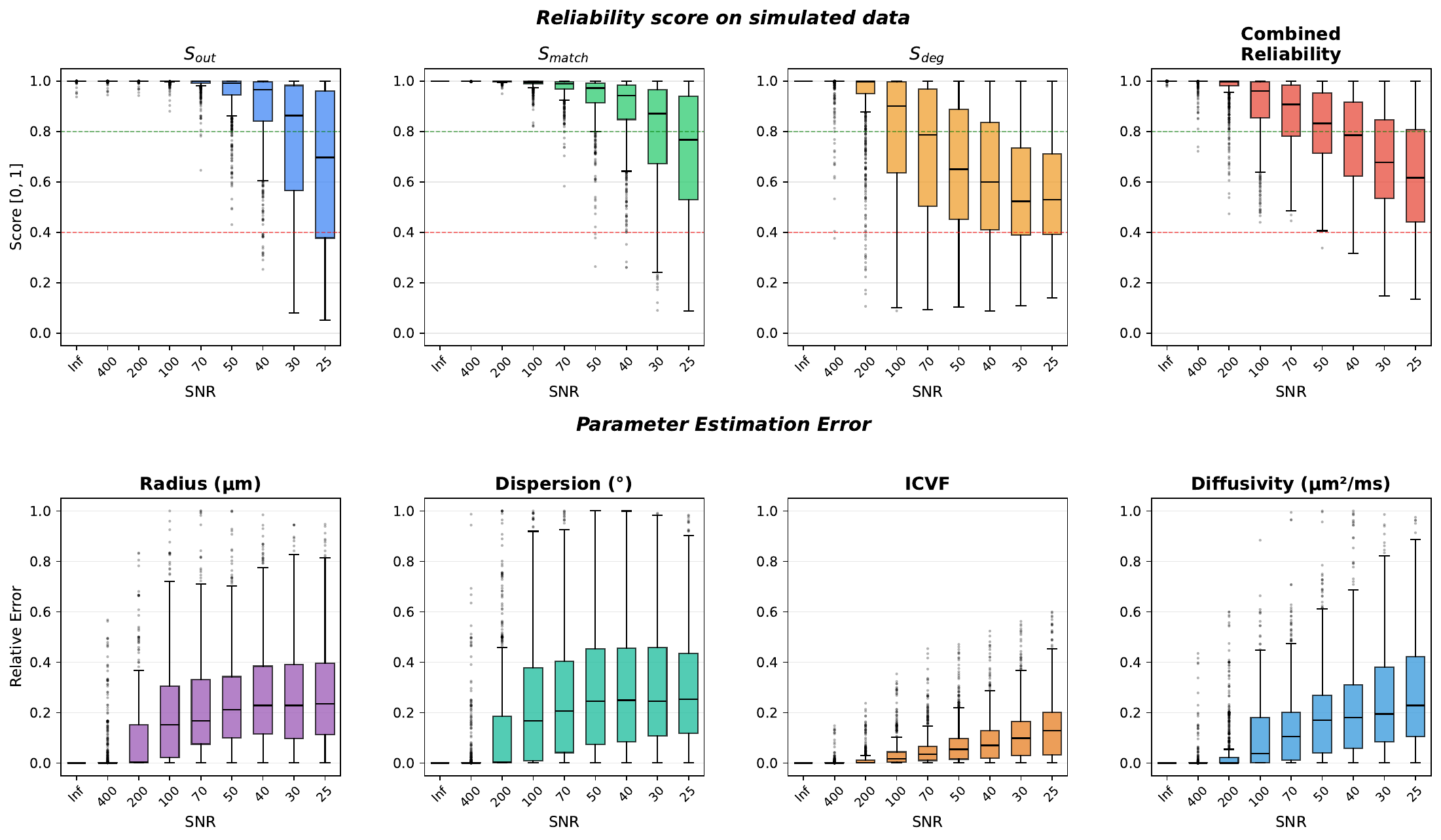}
  \caption{Self-validation of the Reliability Index across SNR levels using leave-one-out testing on the synthetic dictionary. Each column corresponds to an SNR value (horizontal axis, from $\infty$ to~25), and box plots summarise the distribution across 1{,}050 test voxels. \emph{Top row:} Reliability scores $S_{\text{out}}$, $S_{\text{match}}$, $S_{\text{deg}}$, and combined Reliability Index. All scores decrease progressively with increasing noise, with $S_{\text{deg}}$ showing increased variability at higher SNR levels than the other scores. \emph{Bottom row:} Relative estimation error for axon radius, microscopic angular spread ($\mu\theta$), ICVF, and intrinsic diffusivity, normalised by parameter range. ICVF remains the most stable parameter across noise levels, whereas microscopic angular spread shows consistently higher estimation error and variability.}

  \label{fig:reliability_scatter}
\end{figure}

The leave-one-out validation across 1{,}050 synthetic voxels and 9 SNR levels (9{,}450 test cases) showed progressive degradation of all three reliability scores with increasing noise (Figure~\ref{fig:reliability_scatter}). At infinite SNR, most voxels were classified within the reliable tier, whereas lower SNR levels shifted the Reliability Index distributions toward moderate and low reliability ranges. Among the three scores, $S_{\text{deg}}$ showed increased variability at higher SNR levels than $S_{\text{out}}$ and $S_{\text{match}}$, indicating that parameter degeneracy becomes detectable before noise substantially affects signal matching or dictionary support.

Parameter estimation accuracy differed across microstructural parameters (Figure~\ref{fig:reliability_scatter}, bottom row). ICVF showed the lowest estimation error across all SNR levels, remaining below 10\% mean relative error even at SNR~$= 25$. Axon radius and intrinsic diffusivity were accurately recovered at high SNR and degraded progressively with increasing noise, reaching approximately 30\% and 25\% mean relative error at SNR~$= 25$, respectively. Microscopic angular spread~($\mu\theta$) showed consistently higher estimation error, with a mean relative error of approximately 22\% across all SNR levels (IQR: 17-43\%). This elevated error persisted even at high SNR and was accompanied by increased variability in $S_{\text{deg}}$ beginning around SNR~$= 100$ (IQR: 77-97\%), consistent with the intrinsic degeneracy of the two-compartment model\cite{Jelescu2016,Novikov2018,Howard2022}.

The Reliability Index showed a strong negative correlation with mean relative parameter error across all test cases (Spearman $\rho = -0.742$, $p < 10^{-10}$), indicating that lower reliability scores were associated with larger estimation errors.

% --- Figure 2: Reliability scores and estimation error across SNR ---

Supplementary Figures~\ref{fig:synth_high_R}-\ref{fig:synth_low_Sdeg} provide representative synthetic examples associated with each reliability score and source of unreliability.

% ----------------------------------------------------------------------------
\subsection*{Reliability Index and parameter estimates in rat corpus callosum in vivo}
% ----------------------------------------------------------------------------

\begin{figure}[H]
  \centering
  \includegraphics[width=\linewidth]{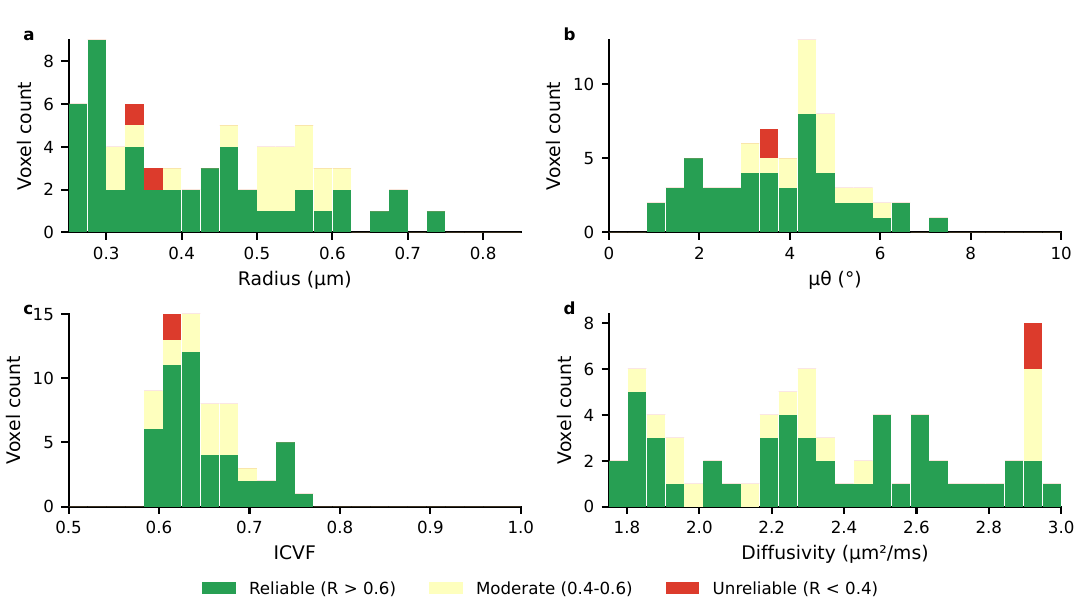}
\caption{Reliability-layered parameter histograms for Rat~1 ($n = 66$ corpus callosum voxels). Each panel shows the voxel distribution of one estimated parameter (radius, microscopic angular spread $\mu\theta$, ICVF, intrinsic diffusivity), with every bar stacked by reliability tier: reliable ($R > 0.6$, green) at the base, moderate ($0.4 \leq R \leq 0.6$, amber), and unreliable ($R < 0.4$, red) on top. The colour stack shows where in parameter space the reliable versus unreliable estimates fall. For Rat~1, 91\% of voxels reach $R > 0.5$. Rats~2 and~4 are shown in Supplementary Figures~\ref{fig:supp_reliability_profile_Vol2}-\ref{fig:supp_reliability_profile_Vol4}.
}
  \label{fig:reliability_profile_Vol1}
\end{figure}

\begin{figure}[H]
  \centering
  \includegraphics[width=\linewidth]{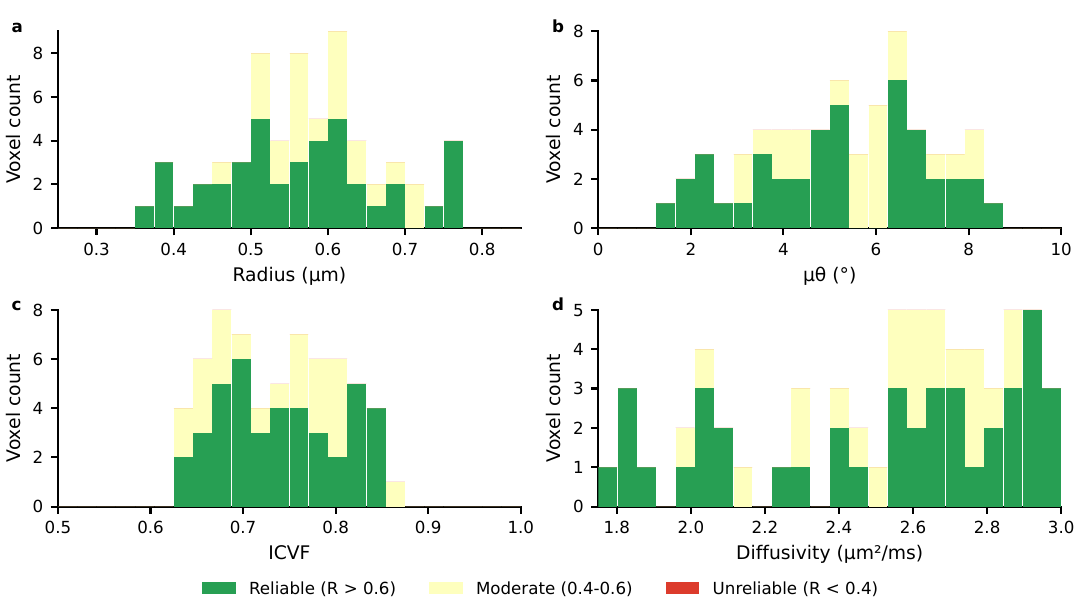}
\caption{Reliability-layered parameter histograms for Rat~3 ($n = 63$ corpus callosum voxels), same format as Figure~\ref{fig:reliability_profile_Vol1}. Rat~3 is the most reliable animal (97\% of voxels with $R > 0.5$) and shows higher median ICVF than the other animals.
}
  \label{fig:reliability_profile_Vol3}
\end{figure}

We next applied the estimation and reliability framework to 256 corpus callosum voxels from four Wistar rats acquired with the NEXI PGSE protocol. Figures~\ref{fig:reliability_profile_Vol1} and~\ref{fig:reliability_profile_Vol3} show, for Rats~1 and~3, the distribution of each estimated microstructural parameter across corpus callosum voxels, with every histogram bar stacked by reliability tier (Rats~2 and~4 are shown in Supplementary Figures~\ref{fig:supp_reliability_profile_Vol2}-\ref{fig:supp_reliability_profile_Vol4}). Reliable voxels ($R > 0.6$) make up the majority of estimates in every animal, and the reliability layering shows where within each parameter distribution the less-reliable estimates fall.

Across all four animals, $S_{\text{out}}$ remained consistently high (median 87-90\%), consistent with substantial overlap between the in vivo corpus callosum signals and the simulated dictionary. $S_{\text{match}}$ was also generally high (median 94-99\% in Rats~1-3 and 68\% in Rat~4), whereas $S_{\text{deg}}$ showed lower values (median 31-41\% in Rats~1-3 and 52\% in Rat~4). Overall, 91\% of corpus callosum voxels exceeded $R > 0.5$ (Table~\ref{tab:per_volume}). $S_{\text{deg}}$ was the lowest of the three scores in all four animals.

Estimated microstructural parameters varied across animals but remained within ranges reported for rat corpus callosum in previous histological and diffusion MRI studies (Table~\ref{tab:per_volume}). Median axon radii ranged from 0.34 to 0.57~$\mu$m, spanning the histological mean radius of approximately 0.35~$\mu$m reported by Pesaresi \emph{et al.}\cite{pesaresi2015axon}. Estimated ICVF values ranged from 0.60 to 0.74, within the intracellular volume fraction of living brain tissue (0.60-0.95)\cite{Tonnesen2018}. Microscopic angular spread estimates ranged from 3.3$^\circ$ to 5.3$^\circ$, and intrinsic diffusivities ranged from 2.08 to 2.70~$\mu$m$^2$/ms, encompassing the approximately 2.25~$\mu$m$^2$/ms values previously reported using planar diffusion encoding\cite{dhital2019intraaxonal}. Rats~1 and~2 showed similar parameter estimates, whereas Rat~3 exhibited larger radius and ICVF estimates.

% --- Table: Per-volume estimates (auto-numbered; renders as Table~\ref{tab:per_volume}) ---
\begin{table}[ht]
\centering
\small
\renewcommand{\arraystretch}{1.3}
\caption{Per-animal microstructural parameter estimates and reliability scores for in vivo rat corpus callosum. Values are medians across voxels. Estimates use exponentially weighted KNN regression ($K = 10$, $\alpha = 10$).}
\label{tab:per_volume}
\begin{tabular}{@{}lccccccccc@{}}
\toprule
Animal & Voxels & Radius & $\mu\theta$ & ICVF & Diffusivity
       & $S_{\text{out}}$ & $S_{\text{match}}$ & $S_{\text{deg}}$
       & RI \\
       & & ($\mu$m) & ($^\circ$) & & ($\mu$m$^2$/ms)
       & (\%) & (\%) & (\%) & (\%) \\
\midrule
Rat\,1 & 66 & 0.42 & 4.1 & 0.64 & 2.32 & 90\% & 96\% & 41\% & 68\% \\
Rat\,2 & 89 & 0.42 & 4.2 & 0.65 & 2.08 & 89\% & 94\% & 38\% & 65\% \\
Rat\,3 & 63 & 0.57 & 5.3 & 0.74 & 2.62 & 89\% & 99\% & 31\% & 62\% \\
Rat\,4 & 38 & 0.34 & 3.3 & 0.60 & 2.70 & 87\% & 68\% & 52\% & 68\% \\
\bottomrule
\end{tabular}
\end{table}

% --- Figure 7: Reliability profile Vol1 ---

% --- Figure 8: Reliability profile Vol3 ---

\begin{figure}[H]
  \centering
  \includegraphics[width=\linewidth]{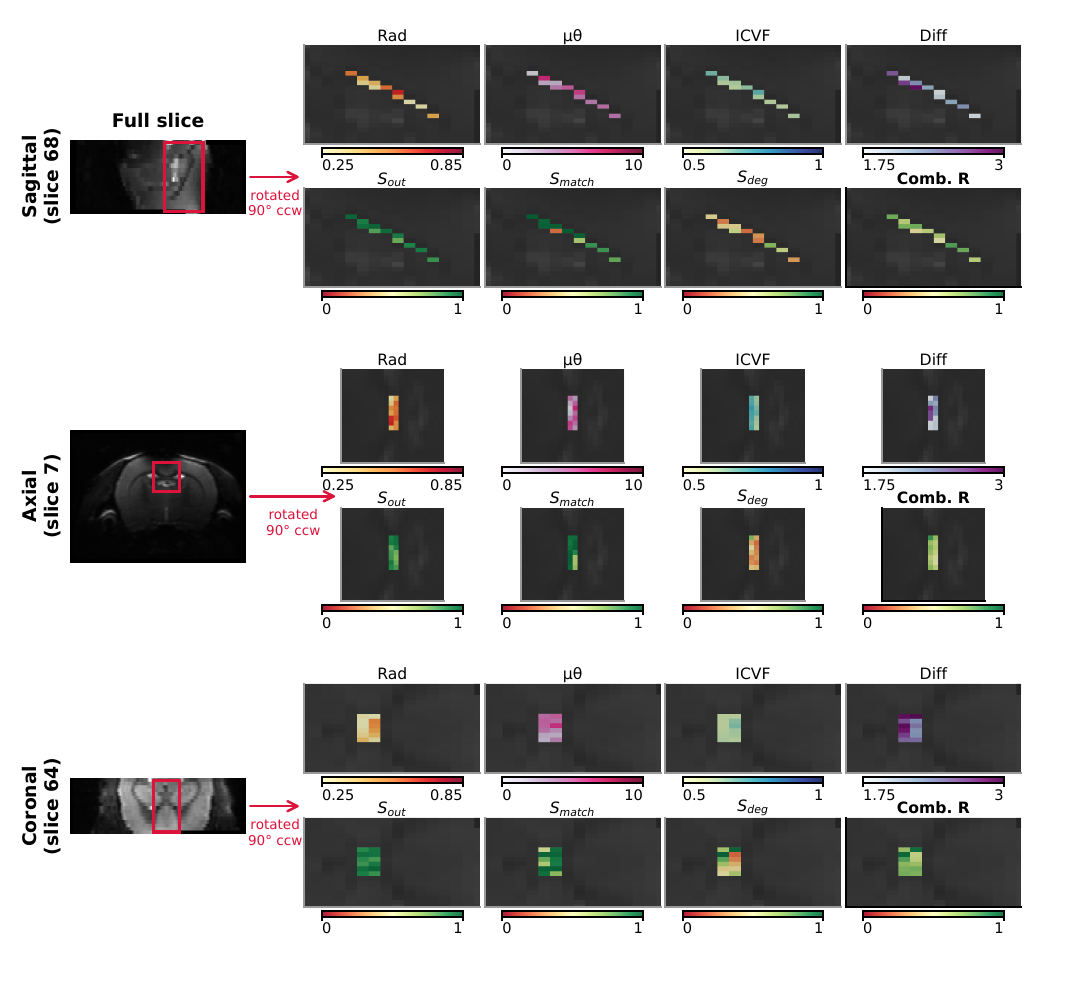}
  \caption{Spatial parameter maps for Rat~1 ($n = 66$ corpus callosum voxels) overlaid on the mean $b_0$ image. Each block shows one projection through the corpus callosum mask (sagittal, axial, then coronal). Within a block, the top row of four panels displays the voxel-wise parameter estimates (radius, microscopic angular spread, ICVF, diffusivity) and the bottom row shows the three reliability scores ($S_{\text{out}}$, $S_{\text{match}}$, $S_{\text{deg}}$) and the combined Reliability Index. Reliability values are generally higher in central corpus callosum regions and lower near mask boundaries, whereas $S_{\text{out}}$ remains uniformly high across most voxels. Rats~2 and~4 are shown in Supplementary Figures~\ref{fig:supp_spatial_maps_Vol2}-\ref{fig:supp_spatial_maps_Vol4}.
}
    \label{fig:spatial_maps_Vol1}
\end{figure}

Figure~\ref{fig:spatial_maps_Vol1} shows the spatial distribution of parameter estimates and reliability scores for Rat~1 (Rats~2 to~4 are shown in Supplementary Figures~\ref{fig:supp_spatial_maps_Vol2}-\ref{fig:supp_spatial_maps_Vol4}). Parameter estimates exhibit spatial continuity across neighbouring corpus callosum voxels. Higher Reliability Index values are predominantly observed in central regions of the mask, whereas lower-reliability voxels are more frequent near mask boundaries. Across animals, $S_{\text{out}}$ remains uniformly high throughout most of the corpus callosum mask.

% --- Figure 4: Spatial parameter maps Volume 1 ---

\begin{figure}[H]
  \centering
  \includegraphics[width=\linewidth]{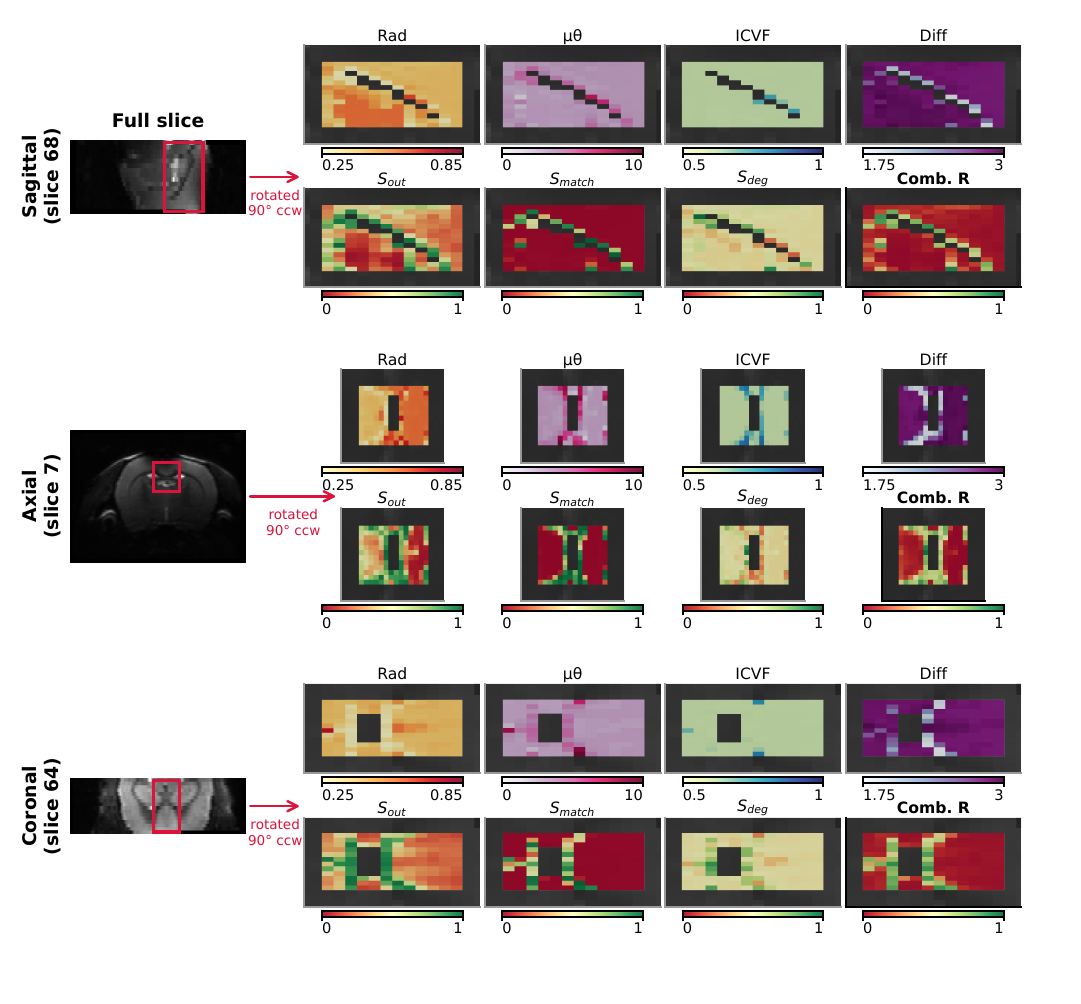}
\caption{Complement spatial parameter maps for Rat~1 overlaid on the same $b_0$ image as Figure~\ref{fig:spatial_maps_Vol1}. The complement region consists of voxels surrounding the corpus callosum mask and excludes the corpus callosum itself. Each block shows one projection (sagittal, axial, then coronal). Within a block, the top row of four panels displays the voxel-wise parameter estimates and the bottom row shows the three reliability scores ($S_{\text{out}}$, $S_{\text{match}}$, $S_{\text{deg}}$) and the combined Reliability Index, using the same colour scales as Figure~\ref{fig:spatial_maps_Vol1}. Relative to the corpus callosum voxels, complement voxels show lower Reliability Index values and reduced $S_{\text{out}}$ scores, while several parameter estimates approach the bounds of the training dictionary. Complement maps for Rats~2-4 are shown in Supplementary Figures~\ref{fig:supp_complement_maps_Vol2}-\ref{fig:supp_complement_maps_Vol4}.
}
  \label{fig:complement_maps_Vol1}
\end{figure}

Applying the estimation framework to the complement region surrounding the corpus callosum produced substantially lower reliability scores than within the corpus callosum mask (Figure~\ref{fig:complement_maps_Vol1}). In particular, $S_{\text{out}}$ decreased markedly across the complement voxels, whereas $S_{\text{match}}$ and the combined Reliability Index also showed reduced values relative to the corpus callosum region. Parameter estimates in the complement region frequently approached the bounds of the training dictionary, with elevated radius and diffusivity estimates and ICVF values concentrated near the edges of the simulated parameter range.

% --- Figure 11: Complement spatial parameter maps Volume 1 ---

% ----------------------------------------------------------------------------
\subsection*{Voxel-level examples on in vivo data}
% ----------------------------------------------------------------------------

\begin{figure}[H]
  \centering
  \includegraphics[width=\linewidth]{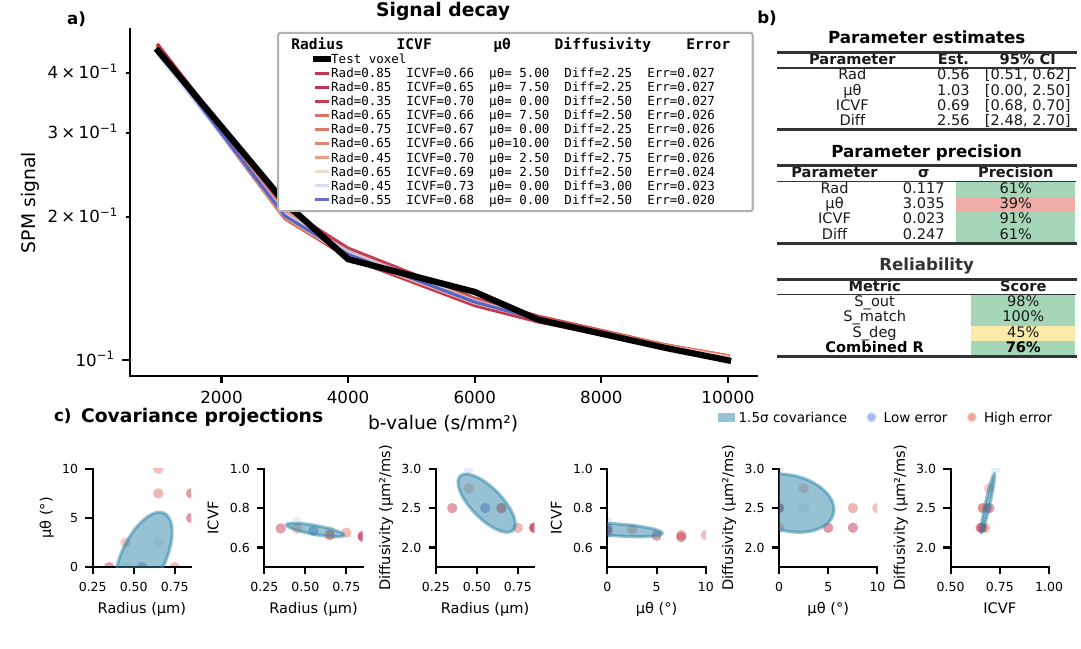}
  \caption{In vivo well-matched example (Rat~1, voxel index~18, $R = 76$\%). \textbf{(a)}~Spherical-mean signal decay (log scale); measured voxel in black, $K = 10$ neighbours coloured by signal distance (blue nearest, red farthest), with a table of their parameters and matching error. \textbf{(b)}~Parameter estimates with 95\% confidence intervals, per-parameter precision $P_j$, and the three reliability scores; precision and score cells shaded green, amber or red by tier. \textbf{(c)}~Pairwise covariance projections with the $1.5\sigma$ ellipse. The signal overlaps its neighbours, and both dictionary support and reconstruction are strong ($S_{\text{out}} = 98$\%, $S_{\text{match}} = 100$\%), while the moderate $S_{\text{deg}} = 45$\% reflects residual parameter spread.}
  \label{fig:invivo_high_R}
\end{figure}

\begin{figure}[H]
  \centering
  \includegraphics[width=\linewidth]{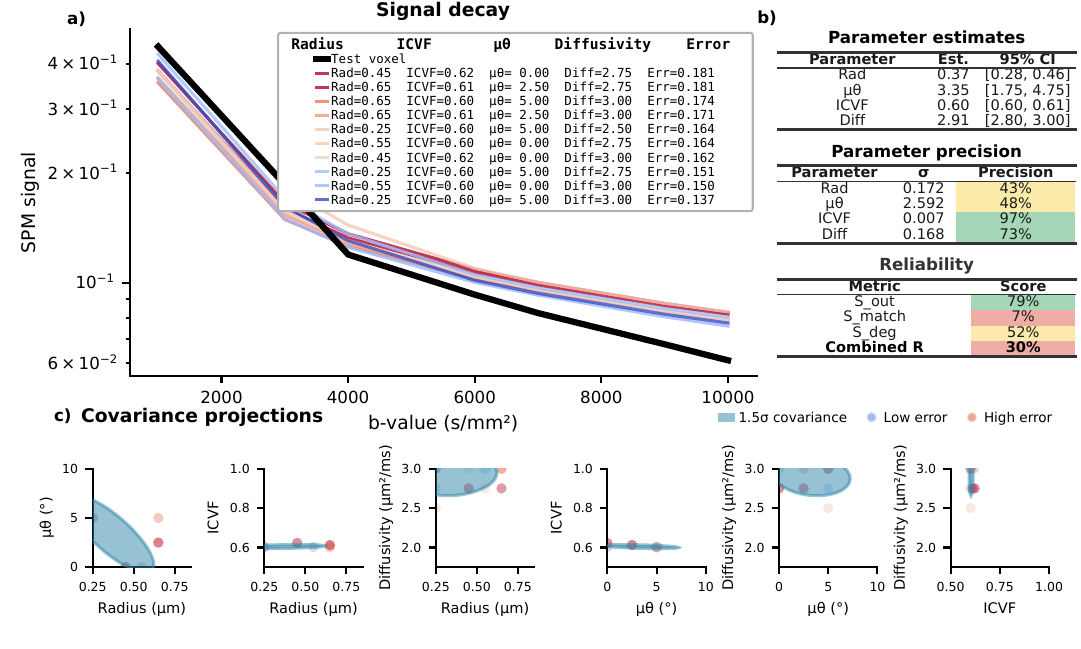}
\caption{Low-$S_{\text{match}}$ in vivo example (Rat~1, voxel index~47, $R = 27$\%). The signal lies within the dictionary support ($S_{\text{out}} = 69$\%), but the nearest neighbours cannot reproduce its shape ($S_{\text{match}} = 4$\%), indicating limited local interpolation or model mismatch; the tightly clustered neighbours give $S_{\text{deg}} = 67$\%. Panels as in Figure~\ref{fig:invivo_high_R}.}
  \label{fig:invivo_low_Smatch}
\end{figure}

\begin{figure}[H]
  \centering
  \includegraphics[width=\linewidth]{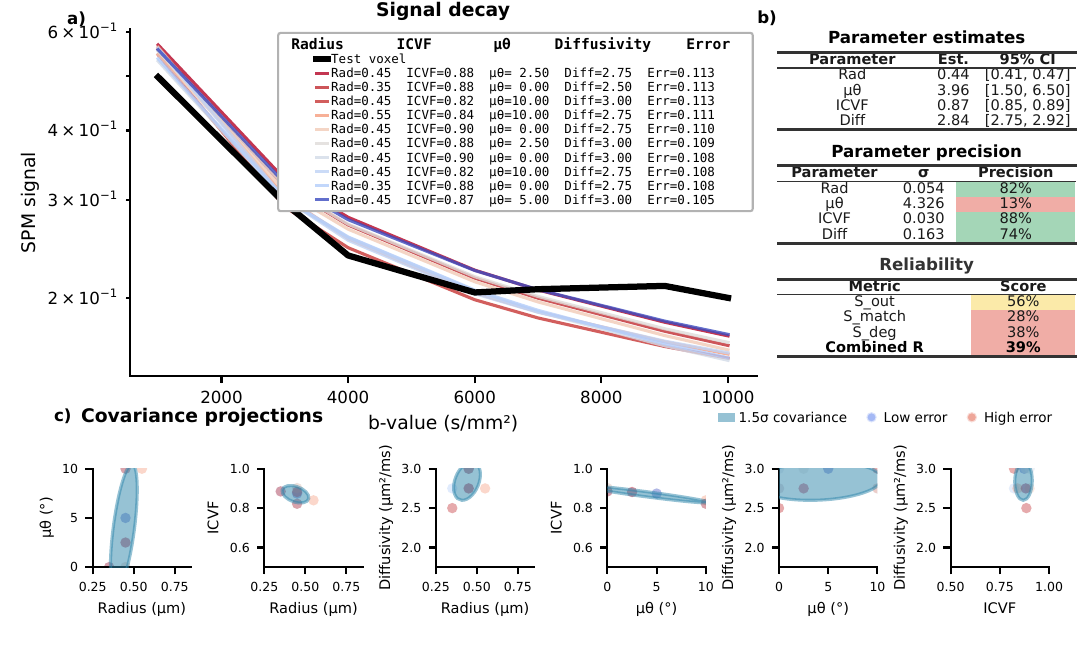}
\caption{In vivo example from outside the corpus callosum that is beginning to leave the dictionary support (Rat~2, complement voxel at $(i,j,k) = (54, 55, 14)$, $R = 39$\%). $S_{\text{out}}$ has fallen to a moderate value ($S_{\text{out}} = 56$\%), consistent with a signal beginning to move outside the dictionary support, while the nearest neighbours reconstruct it poorly ($S_{\text{match}} = 28$\%) and $S_{\text{deg}} = 38$\%. The combined Reliability Index ($R = 39$\%) falls in the unreliable range. Panels as in Figure~\ref{fig:invivo_high_R}.}
  \label{fig:invivo_low_Sout}
\end{figure}

Figures~\ref{fig:invivo_high_R}-\ref{fig:invivo_low_Sout} show representative in vivo voxel examples spanning the range of reliability outcomes. Each figure has three panels: \textbf{(a)} the measured spherical-mean signal decay (black) with its $K = 10$ nearest dictionary neighbours, coloured from blue (nearest) to red (farthest in signal distance), and a table of each neighbour's parameters and matching error; \textbf{(b)} the parameter estimates with bootstrap 95\% confidence intervals, the per-parameter precision $P_j$ (Eq.~\ref{eq:precision}), and the three reliability scores with the Reliability Index; and \textbf{(c)} the pairwise covariance projections of the retained neighbours with the $1.5\sigma$ ellipse. In panel~(b), the precision and reliability cells are shaded green (well constrained or reliable) through amber to red (poorly constrained or unreliable) using the reliability-tier thresholds. Across cases, radius and microscopic angular spread estimates exhibit a consistent negative covariance structure.

% --- Figure 12: In-vivo well-matched example (label kept; content repurposed to Rat1 v18) ---

% --- Figure: In-vivo low S_match (Rat1 v47); shown before the OOD example so the "signal still in the dictionary family" case precedes the "signal outside the dictionary" case ---

% --- Figure: In-vivo low S_out (OOD complement voxel) ---

Together, the three examples span the range of reliability outcomes, from a well-constrained voxel to a voxel the neighbours cannot reconstruct (low $S_{\text{match}}$) and a voxel beginning to leave the dictionary support (reduced $S_{\text{out}}$).

% ----------------------------------------------------------------------------
\subsection*{Reliability Index in human corpus callosum}
% ----------------------------------------------------------------------------

\begin{figure}[H]
  \centering
  \includegraphics[width=\linewidth]{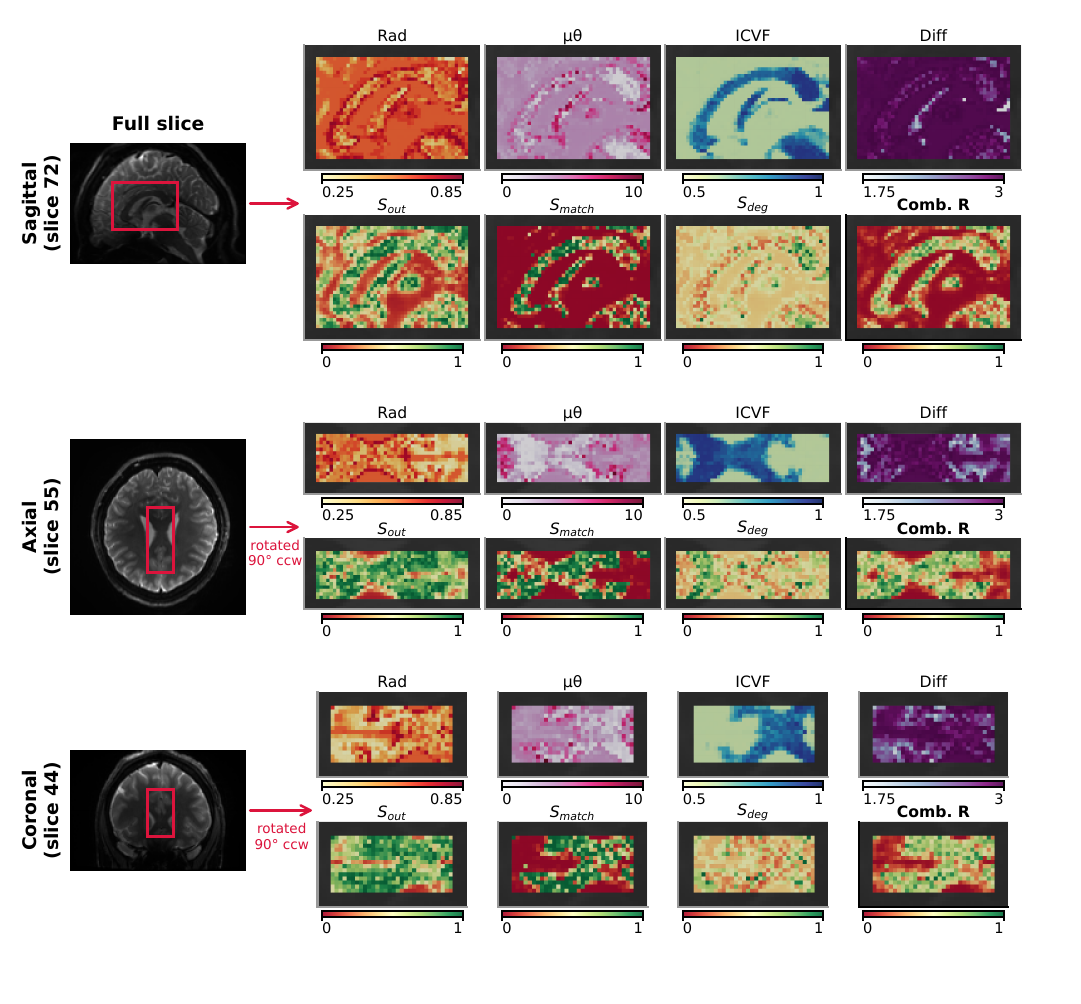}
  \caption{Spatial parameter and reliability maps for a representative human subject (Subject~3), shown over the union of the corpus callosum and its surrounding complement region ($\Omega$). Each block shows one projection (sagittal, axial, then coronal); within a block, the top row of panels displays the voxel-wise parameter estimates (radius, microscopic angular spread, ICVF, diffusivity) and the bottom row shows the three reliability scores ($S_{\text{out}}$, $S_{\text{match}}$, $S_{\text{deg}}$) and the combined Reliability Index. The corpus callosum stands out as a high-reliability core within a low-reliability surround, mirroring the rat maps.}
  \label{fig:hcp_omega}
\end{figure}

\begin{figure}[H]
  \centering
  \includegraphics[width=\linewidth]{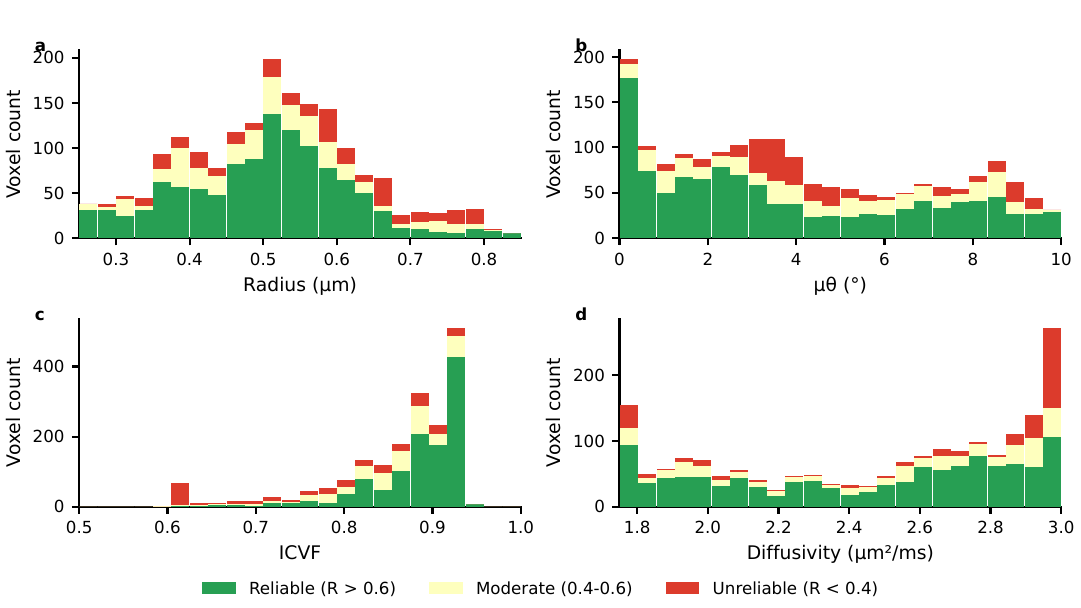}
  \caption{Reliability-layered parameter histograms for Subject~4, same format as Figure~\ref{fig:reliability_profile_Vol1}. 76\% of voxels reach $R > 0.5$. }
  \label{fig:hcp_hist}
\end{figure}

We applied the same estimation and reliability framework to 18{,}765 corpus callosum voxels from nine subjects of the MGH-USC Human Connectome Project dataset\cite{Setsompop2013,Fan2016} (Table~\ref{tab:per_subject_hcp}). Across subjects, 73\% of corpus callosum voxels exceeded $R > 0.5$ (55 to 77\% per subject), with per-subject median Reliability Index between 50\% and 65\%. Reliability was again concentrated in the central corpus callosum, with $S_{\text{out}}$ and $S_{\text{match}}$ high across the tract (Figures~\ref{fig:hcp_omega} and~\ref{fig:hcp_hist}). As in the rat dataset, $S_{\text{deg}}$ was the lowest of the three scores in most subjects, whereas Subject~3 was limited by $S_{\text{match}}$ (median 42\%). The reliability-layered histograms show the same three-tier structure as in rat, with reliable estimates ($R > 0.6$) forming the majority in most subjects.

% --- Table: Per-subject HCP estimates (new) ---
\begin{table}[ht]
\centering
\small
\renewcommand{\arraystretch}{1.3}
\caption{Per-subject microstructural parameter estimates and reliability scores for in vivo human corpus callosum (MGH-USC Human Connectome Project dataset). Values are medians across voxels. Estimates use exponentially weighted KNN regression ($K = 10$, $\alpha = 10$).}
\label{tab:per_subject_hcp}
\begin{tabular}{@{}lccccccccc@{}}
\toprule
Subject & Voxels & Radius & $\mu\theta$ & ICVF & Diffusivity
       & $S_{\text{out}}$ & $S_{\text{match}}$ & $S_{\text{deg}}$
       & RI \\
       & & ($\mu$m) & ($^\circ$) & & ($\mu$m$^2$/ms)
       & (\%) & (\%) & (\%) & (\%) \\
\midrule
Subject~1 & 1{,}580 & 0.53 & 7.0 & 0.88 & 2.22 & 85\% & 89\% & 37\% & 62\% \\
Subject~2 & 2{,}374 & 0.58 & 3.1 & 0.89 & 2.65 & 85\% & 89\% & 48\% & 65\% \\
Subject~3 & 1{,}795 & 0.62 & 2.7 & 0.85 & 2.91 & 71\% & 42\% & 47\% & 50\% \\
Subject~4 & 1{,}837 & 0.57 & 3.5 & 0.89 & 2.60 & 86\% & 91\% & 49\% & 65\% \\
Subject~5 & 1{,}764 & 0.62 & 2.3 & 0.85 & 2.85 & 85\% & 86\% & 46\% & 63\% \\
Subject~6 & 1{,}988 & 0.58 & 2.2 & 0.89 & 2.74 & 84\% & 86\% & 49\% & 64\% \\
Subject~7 & 2{,}580 & 0.61 & 2.4 & 0.85 & 2.84 & 82\% & 78\% & 46\% & 61\% \\
Subject~8 & 2{,}317 & 0.56 & 3.6 & 0.89 & 2.38 & 86\% & 92\% & 49\% & 65\% \\
Subject~9 & 2{,}530 & 0.61 & 2.5 & 0.85 & 2.86 & 81\% & 79\% & 42\% & 61\% \\
\bottomrule
\end{tabular}
\end{table}
\clearpage   % not \newpage

% --- Figure: HCP omega spatial map (MGH1003) ---

% --- Figure: HCP reliability-layered histogram (MGH1004) ---

% --- Figure: HCP high-reliability voxel (MGH1001 v1168) ---
\begin{figure}[H]
  \centering
  \includegraphics[width=\linewidth]{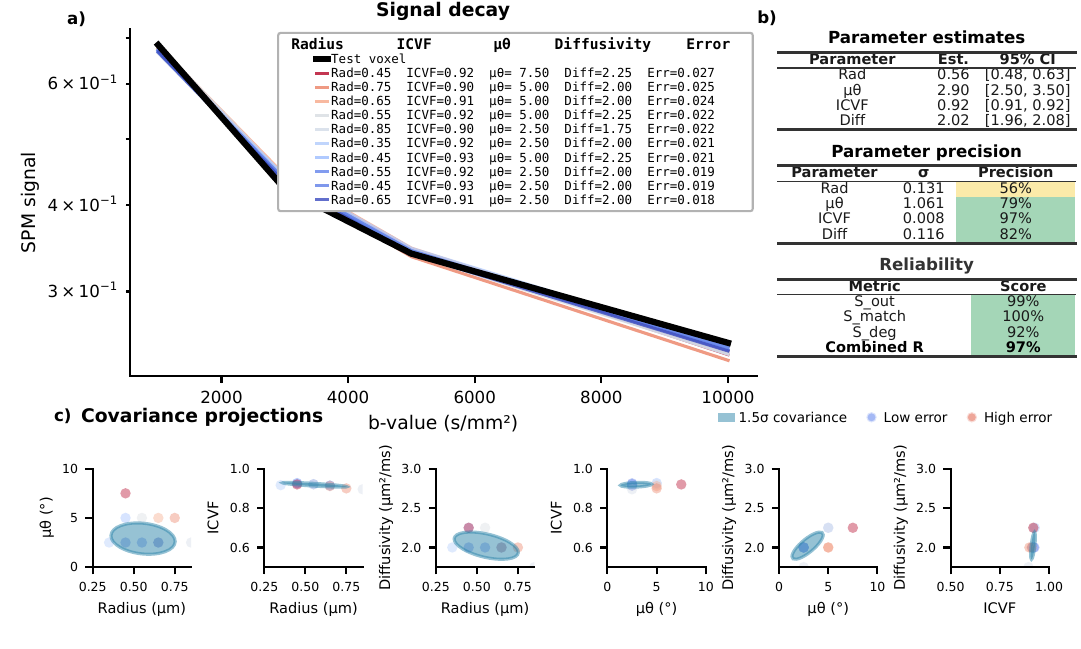}
  \caption{High-reliability human example (Subject~1, voxel index~1168, $R = 97$\%). All three pillars are high ($S_{\text{out}} = 99$\%, $S_{\text{match}} = 100$\%, $S_{\text{deg}} = 92$\%): the signal is well inside the dictionary, accurately reconstructed, and the retained neighbours pin the parameters tightly. Panels as in Figure~\ref{fig:invivo_high_R}.}
  \label{fig:hcp_good}
\end{figure}

% --- Figure: HCP high-precision but unreliable voxel (MGH1001 v846) ---
\begin{figure}[H]
  \centering
  \includegraphics[width=\linewidth]{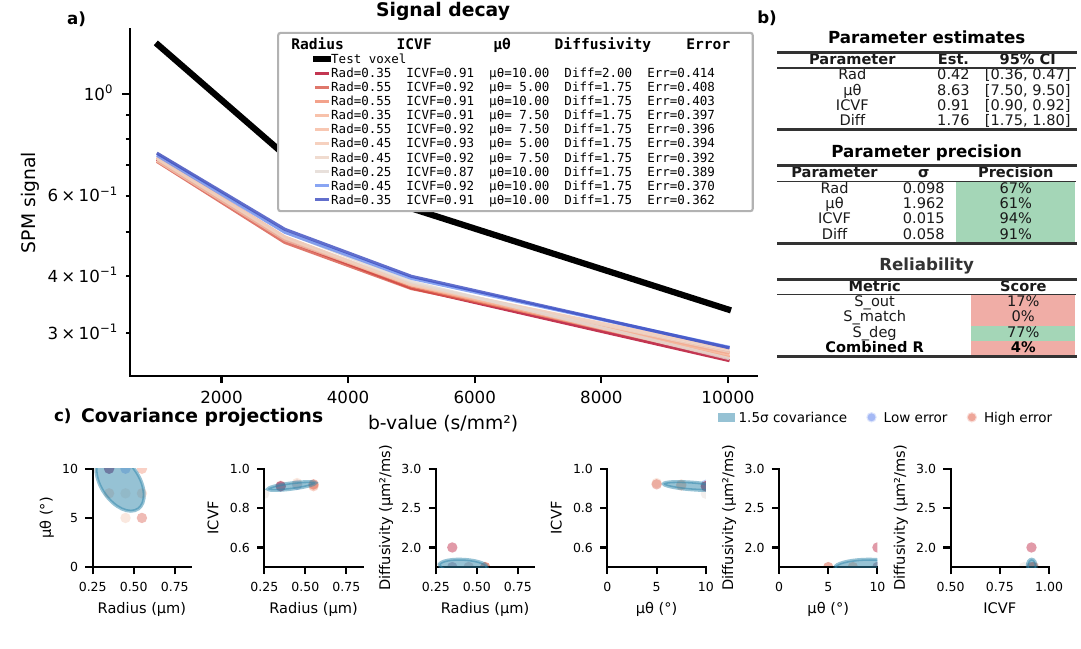}
  \caption{Human example where per-parameter precision does not imply reliability (Subject~1, voxel index~846, $R = 4$\%). The retained neighbours agree tightly, giving a high degeneracy score ($S_{\text{deg}} = 77$\%), yet the signal lies outside the dictionary support and cannot be reconstructed ($S_{\text{out}} = 17$\%, $S_{\text{match}} = 0$\%). A high $S_{\text{deg}}$ alone is therefore not evidence of a trustworthy estimate; it is informative only once $S_{\text{out}}$ and $S_{\text{match}}$ exceed 60\%. Panels as in Figure~\ref{fig:invivo_high_R}.}
  \label{fig:hcp_precision}
\end{figure}

% discussion.tex — ~1,200 words, NO subheadings

\section*{Discussion and Conclusions}

% We present a framework that connects the three legs of white matter microstructure validation: histological priors ground the dictionary parameter ranges, Monte Carlo simulations on CACTUS substrates translate those priors into a realistic signal space, and in~vivo DW-MRI measurements test whether the simulated dictionary can explain experimental data. Histological electron-microscopy statistics define the substrate parameters (axon radius, packing fraction, dispersion); the high $S_{\text{out}}$ observed across all animals (Table~\ref{tab:per_volume}) confirms that the resulting dictionary covers the in~vivo signal space; and estimated parameters fall within histological reference ranges (Table~\ref{tab:per_volume}). One leg remains incomplete: ideally, histology would come from the same animals, whereas here we rely on published electron-microscopy statistics from the same species and tract\cite{pesaresi2015axon}. The Reliability Index accompanies each estimate, pinpointing which source of uncertainty (dictionary coverage, reconstruction fidelity, or parameter degeneracy) limits confidence.

This work introduces a reliability framework for simulation-based DW-MRI microstructure estimation and demonstrates it using histology-informed Monte Carlo dictionaries of rat corpus callosum, applied to in vivo rat and human corpus callosum. The framework links three elements: parameter ranges derived from electron microscopy, realistic CACTUS-based substrates simulated with MC-DC, and in vivo DW-MRI acquired in the same tissue. Rather than treating dictionary matching as a source of parameter estimates alone, the proposed Reliability Index characterises whether each estimate is limited by insufficient dictionary support, local signal mismatch, or parameter degeneracy. In the present dataset, high $S_{\text{out}}$ values indicate substantial overlap between the selected in vivo corpus callosum signals and the simulated dictionary, while the estimated parameters remain within ranges reported in prior histological and diffusion MRI studies. Same-animal histology was not available, so this comparison supports biological plausibility rather than direct validation.

Leave-one-out validation confirms that the Reliability Index tracks estimation error across noise levels ($\rho = -0.742$, $p < 10^{-10}$). Among the three scores, $S_{\text{deg}}$ exhibited increased variability at higher SNR levels than $S_{\text{out}}$ and $S_{\text{match}}$, indicating that parameter degeneracy emerges even when signal matching and dictionary support remain relatively stable. This behaviour is consistent with the intrinsic degeneracy of the two-compartment model\cite{Jelescu2016,Novikov2018,Howard2022}, which persists even with fully directional acquisitions, rather than with noise alone.

% These results clarify what the 40\% moderate-to-reliable rate means for the field. In well-aligned rat corpus callosum (a favourable tissue with histology-informed priors), 60\% of voxels remain questionable or unreliable. This is not a failure of the method; it reveals the fundamental difficulty of microstructure estimation from DW-MRI. Without reliability scoring, all 65 estimates appear equally trustworthy, and downstream analyses inherit hidden uncertainty. The framework exposes that uncertainty rather than concealing it. The fact that parameter degeneracy ($S_{\text{deg}}$) is the dominant limitation across all animals and SNR levels (Table~\ref{tab:per_volume}) points to a fundamental property of the spherical mean representation, not a shortcoming of the dictionary or estimator. This motivates the adoption of richer signal representations that retain more microstructural contrast.

The in vivo results show that, in well-aligned rat corpus callosum with histology-informed priors, most corpus callosum voxels exceeded $R > 0.5$ (91\%). Across animals, $S_{\text{deg}}$ was consistently lower than $S_{\text{out}}$ and $S_{\text{match}}$, so parameter degeneracy remained the tightest of the three constraints even where the measured signals were well represented in the dictionary. Because the Reliability Index is a geometric mean, this tempers the combined score rather than rejecting the voxel outright. Without voxel-wise reliability assessment, these differences in estimation stability would not be directly observable from the parameter maps alone. Together, these findings motivate the use of richer signal representations that preserve additional microstructural contrast.

A central aspect of the proposed framework is that different sources of unreliability produce distinct score profiles. Unlike global uncertainty measures alone, the proposed decomposition separates uncertainty arising from dictionary coverage, local reconstruction fidelity, and parameter degeneracy. Low $S_{\text{out}}$ values indicate that the measured signal falls outside the region represented by the simulated dictionary, whereas low $S_{\text{match}}$ values reflect insufficient local reconstruction despite adequate signal-space coverage. In contrast, low $S_{\text{deg}}$ values arise when multiple parameter configurations remain compatible with the observed signal. The covariance projections further indicate which parameter combinations contribute most strongly to this ambiguity. $S_{\text{out}}$ and $S_{\text{match}}$ are most useful when read together, since this separates a poor local match from a voxel that is leaving the dictionary. In the complement voxel of Figure~\ref{fig:invivo_low_Sout}, $S_{\text{match}}$ is low but not zero ($28$\%) and $S_{\text{out}}$ is also low ($56$\%). These values indicate that the voxel does not match the dictionary well locally, and that its signal is starting to fall outside the dictionary as well, so the voxel is becoming an outlier rather than only a poor local match. Rather than reducing reliability to a single uncertainty metric, this decomposition helps identify whether limitations originate from dictionary support, local sampling density, or intrinsic parameter degeneracy. In practice, these observations guided iterative refinement of the present framework: early dictionaries with narrower parameterisation produced reduced $S_{\text{out}}$ values, while sparse parameter sampling led to reduced $S_{\text{match}}$ scores and motivated denser dictionary construction.

Simulation-based and dictionary-matching approaches have previously been used to estimate microstructure from diffusion MRI, including convex-optimisation dictionary fitting\cite{Daducci2015}, Monte Carlo signal dictionaries\cite{Rensonnet2019}, and machine-learning estimators trained on simulated signals\cite{Nedjati-Gilani2017}. A parallel line of work quantifies estimation uncertainty, including Bayesian posterior estimation\cite{Reisert2017,Jallais2024} and simulation-based inference\cite{Cranmer2020}, which return point estimates or posterior distributions for a given forward model. The present framework adds a complementary per-voxel decomposition that separates whether an estimate is limited by dictionary coverage, local reconstruction, or parameter degeneracy, relative to the simulated dictionary and acquisition protocol.

The human dataset followed the same pattern as the rat data. Reliability was concentrated in the central corpus callosum, with $S_{\text{out}}$ and $S_{\text{match}}$ high across the tract and $S_{\text{deg}}$ again the lowest of the three scores, so that parameter degeneracy rather than dictionary support set the ceiling on reliability in most subjects (Figures~\ref{fig:hcp_omega} and~\ref{fig:hcp_hist}); Subject~3 was the exception, limited by $S_{\text{match}}$. The human voxel-level examples reproduced the two informative cases seen in rat: a voxel with all three pillars strong ($R = 97$\%, Figure~\ref{fig:hcp_good}), and one where tightly clustered neighbours produced a high $S_{\text{deg}}$ even though the signal fell outside the dictionary support ($R = 4$\%, Figure~\ref{fig:hcp_precision}), indicating that per-parameter precision is interpretable only once $S_{\text{out}}$ and $S_{\text{match}}$ exceed 60\%. That a rat-derived dictionary transferred to the human corpus callosum without a collapse in $S_{\text{out}}$ is itself informative: the substrates were built from rat histology, whose mean axon radii and microscopic angular spread fall within the ranges reported for human corpus callosum\cite{Liewald2014,Mollink2017}. Any human signal outside the simulated support would have been flagged by a reduced $S_{\text{out}}$. The consistently high $S_{\text{out}}$ therefore indicates that the acquired spherical-mean signals were well represented, so human reliability was again limited by parameter degeneracy rather than by cross-species mismatch. 

The present dictionary was grounded in corpus callosum histology and therefore represents highly aligned white matter with low microscopic angular spread. Extending the framework to whole-brain analysis would require broader parameter ranges and additional tissue configurations, including higher and more heterogeneous microscopic angular spread, crossing fibres, and grey-matter compartments. Such an extension is expected to increase parameter degeneracy, since a wider range of configurations becomes compatible with the measured signal; within the proposed framework this would be reflected by reduced $S_{\text{out}}$ for tissue outside the dictionary and reduced $S_{\text{deg}}$. The reliability decomposition itself remains independent of the underlying substrate model and applies unchanged to such extended dictionaries.

The present framework also suggests natural directions for future development. We chose to powder-average the directional acquisitions into a spherical-mean representation, which erodes the fibre orientation distribution but does not completely discard the microstructural information. The persistent parameter degeneracy observed across animals, particularly for the microscopic angular spread, is therefore not attributable to powder-averaging alone, but is intrinsic to the two-compartment model\cite{Jelescu2016,Novikov2018,Howard2022}. Richer representations that preserve directional contrast, including tensor-valued diffusion encoding\cite{Topgaard2017}, may help improve parameter identifiability. Our simulations omit compartment-specific $T_2$ relaxation, so the fitted weight corresponds to a geometric volume fraction and could be extended to a signal fraction to capture the shorter extra-axonal than intra-axonal $T_2$ reported for white matter\cite{Veraart2018,mckinnon2019measuring}. 
The present study remains a proof-of-concept demonstration based on 256 corpus callosum voxels from four rats and 18{,}765 voxels from nine human subjects, using a simplified two-compartment model that does not explicitly include myelin water or CSF contributions. Extension to other tissues or species will require tissue-specific dictionary generation, although the reliability framework itself remains independent of the underlying substrate model. More broadly, the decomposition of reliability into distinct sources of unreliability naturally supports iterative refinement of simulation-based pipelines. Additional microstructural features such as glial cells, crossing fibres, or pathological morphologies can be incorporated progressively within the same estimation framework, while the simulated dictionaries may also serve as training data for simulation-assisted learning approaches\cite{Rafael-Patino2020a}.

% --- Supplementary material ---
% supplementary.tex — Supplementary Material

\section*{Supplementary Material}

% --- Supplementary numbering: restart counters and prefix labels with "S" ---
\setcounter{figure}{0}
\setcounter{table}{0}
\renewcommand{\thefigure}{S\arabic{figure}}
\renewcommand{\thetable}{S\arabic{table}}

% ---- Synthetic voxel-level examples (moved from main text) ----

% --- Figure S1: Synthetic high R ---
\begin{figure}[H]
  \centering
  \includegraphics[width=\linewidth]{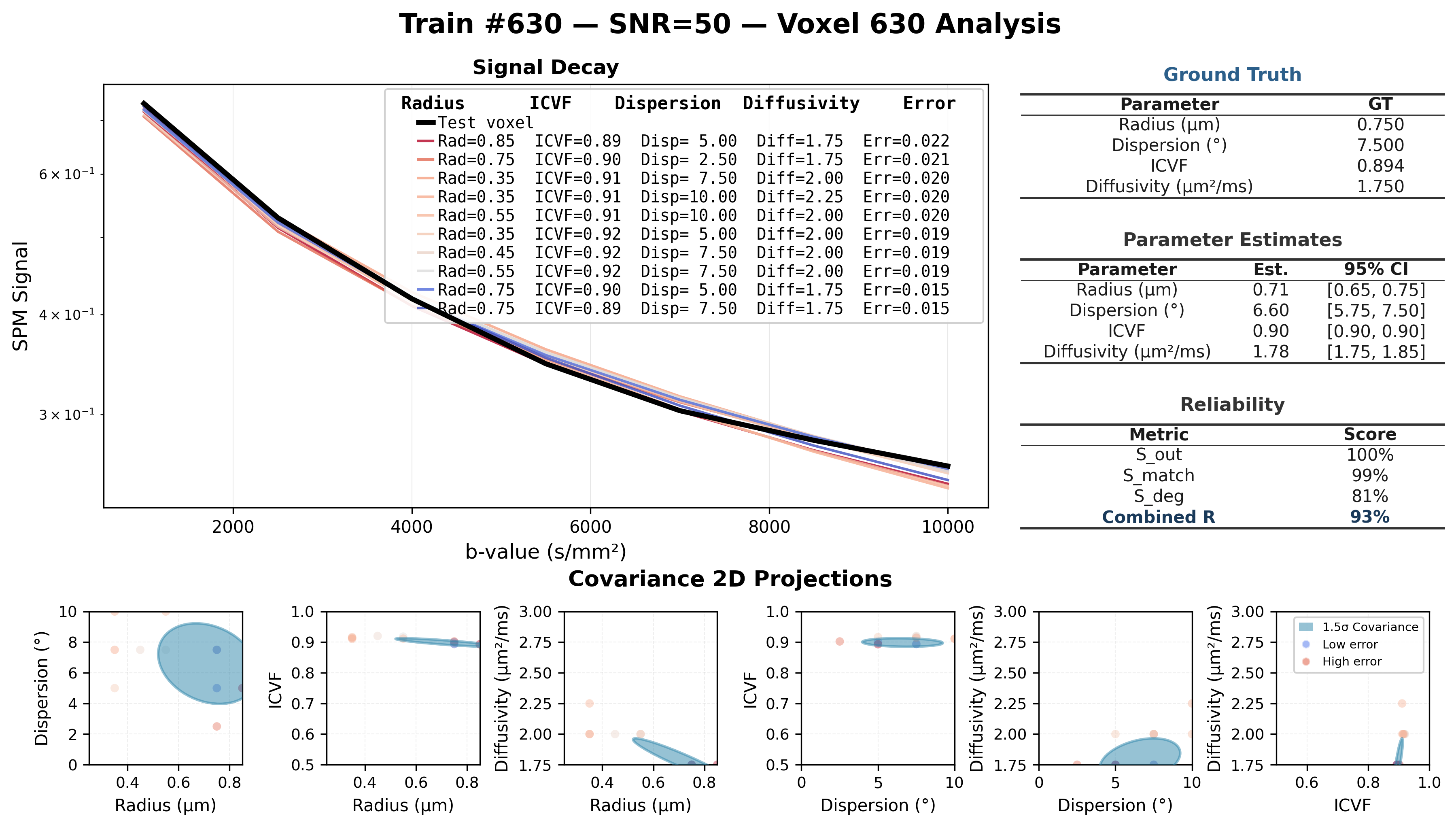}
  \caption{Synthetic high-reliability example (substrate~630, SNR~$= 50$, $R = 93$\%). \emph{Top}: signal decay (log scale) of the evaluated voxel (black line) with its $K = 10$ nearest neighbours colour-coded by distance, alongside ground-truth values, parameter estimates, and reliability scores ($S_{\text{out}} = 100$\%, $S_{\text{match}} = 100$\%, $S_{\text{deg}} = 80$\%). \emph{Bottom}: pairwise covariance 2D projections. All neighbour curves closely overlap the target signal, covariance ellipses are tight, and the estimate is well-constrained.}
  \label{fig:synth_high_R}
\end{figure}

% --- Figure S2: Synthetic low S_out ---
\begin{figure}[H]
  \centering
  \includegraphics[width=\linewidth]{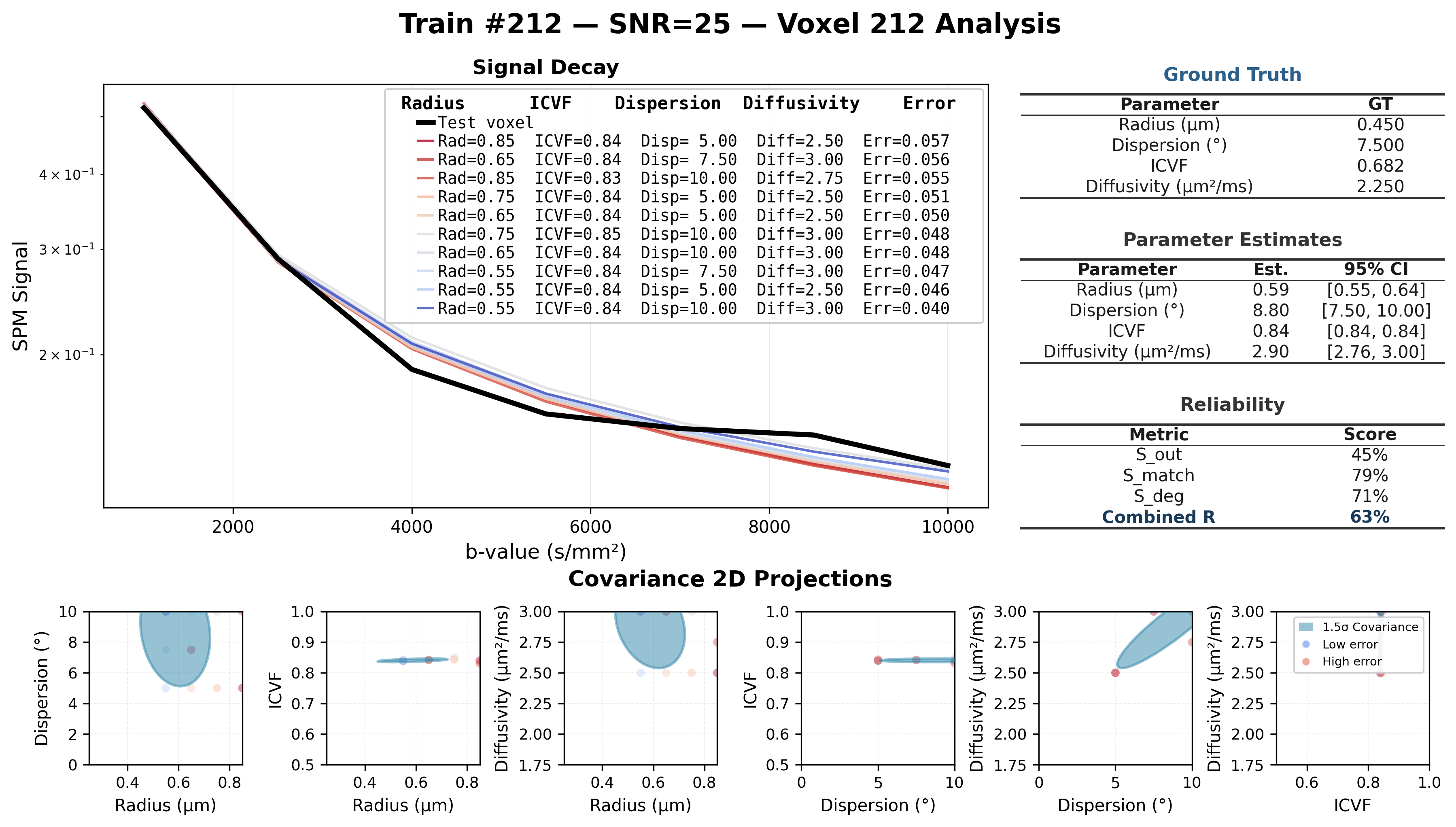}
  \caption{Synthetic low-$S_{\text{out}}$ example (substrate~212, SNR~$= 25$, $R = 52$\%). Same format as Figure~\ref{fig:synth_high_R}. $S_{\text{out}} = 35$\% is the dominant limitation ($S_{\text{match}} = 66$\%, $S_{\text{deg}} = 61$\%): the noisy signal falls near the boundary of the dictionary's signal space. Neighbour curves diverge visibly from the target, particularly at high $b$-values where noise is most pronounced.}
  \label{fig:synth_low_Sout}
\end{figure}

% --- Figure S3: Synthetic low S_match ---
\begin{figure}[H]
  \centering
  \includegraphics[width=\linewidth]{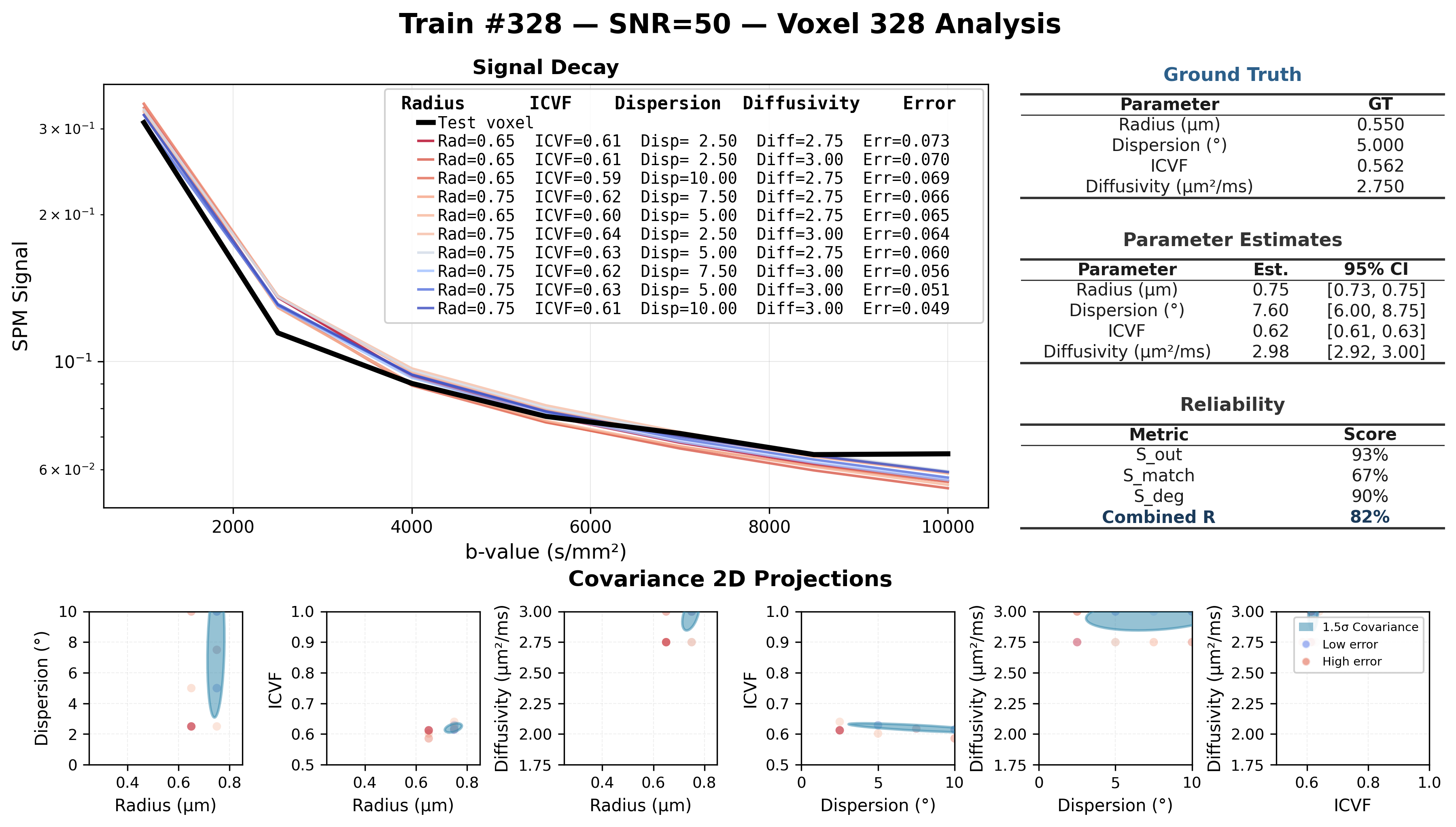}
  \caption{Synthetic low-$S_{\text{match}}$ example (substrate~328, SNR~$= 50$, $R = 54$\%). Same format as Figure~\ref{fig:synth_high_R}. $S_{\text{match}} = 32$\% is the dominant limitation ($S_{\text{out}} = 61$\%, $S_{\text{deg}} = 81$\%): the signal is within the dictionary's support and the parameter estimate is well-constrained, but systematic residuals between the target and the nearest neighbours indicate that the local dictionary density is insufficient to accurately interpolate the target signal.}
  \label{fig:synth_low_Smatch}
\end{figure}

% --- Figure S4: Synthetic low S_deg ---
\begin{figure}[H]
  \centering
  \includegraphics[width=\linewidth]{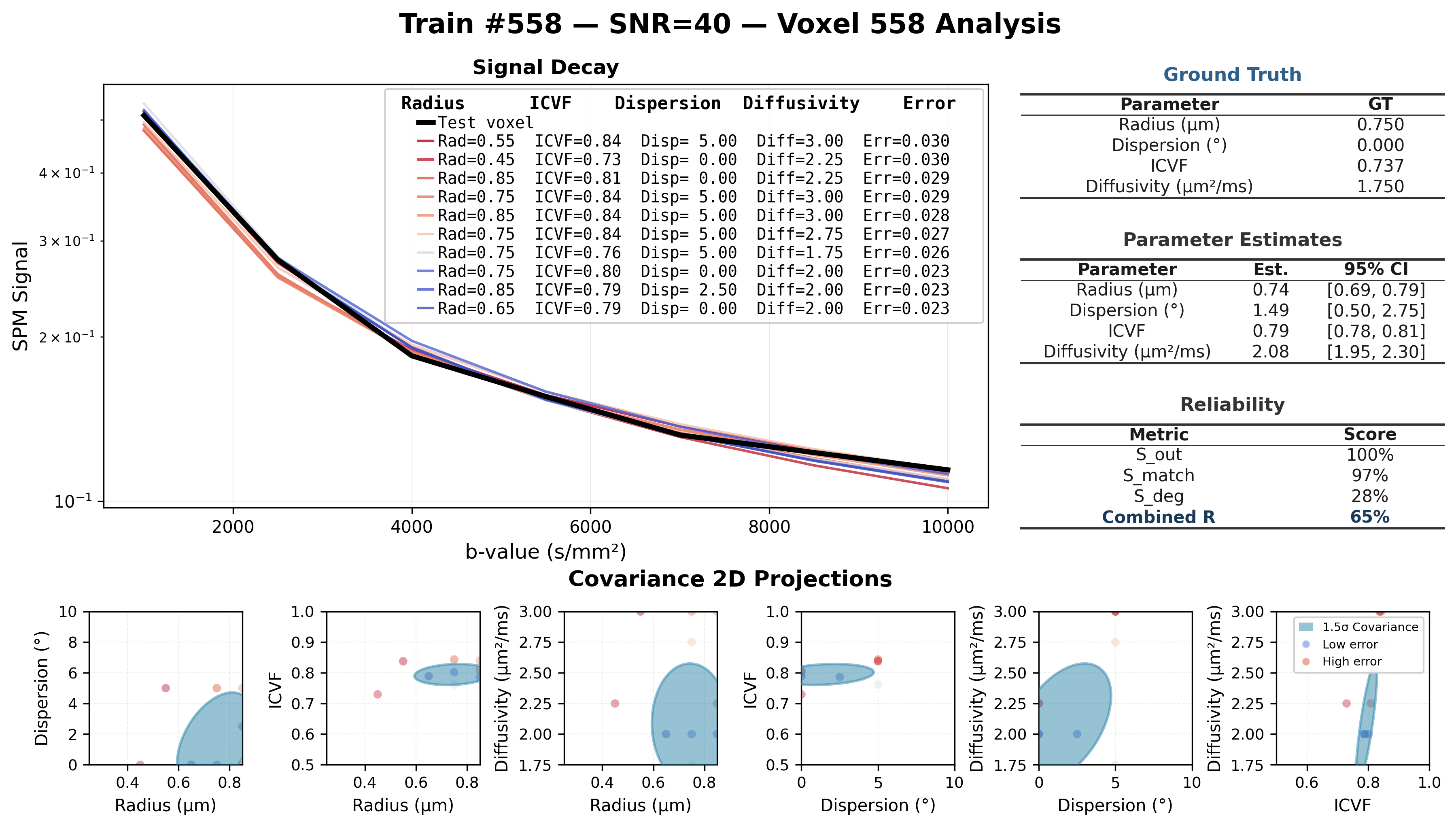}
  \caption{Synthetic low-$S_{\text{deg}}$ example (substrate~558, SNR~$= 40$, $R = 46$\%). Same format as Figure~\ref{fig:synth_high_R}. $S_{\text{deg}} = 12$\% is the dominant limitation ($S_{\text{out}} = 97$\%, $S_{\text{match}} = 86$\%): neighbour curves closely overlap the target signal, but the covariance ellipses are elongated and widely spread, showing that distinct parameter configurations produce nearly identical signals.}
  \label{fig:synth_low_Sdeg}
\end{figure}

% ---- Per-volume supplementary figures ----

% --- Figure S5: Reliability profile Vol2 ---
\begin{figure}[H]
  \centering
  \includegraphics[width=\linewidth]{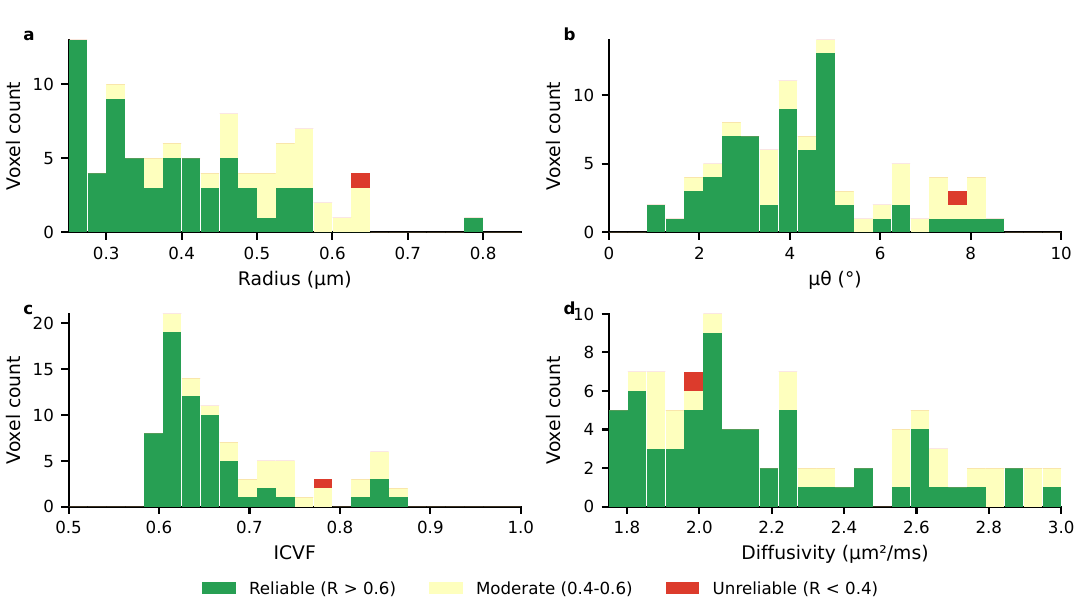}
  \caption{Reliability-layered parameter histograms for Rat~2 ($n = 89$ corpus callosum voxels), same format as Figure~\ref{fig:reliability_profile_Vol1}. 94\% of voxels reach $R > 0.5$.}
  \label{fig:supp_reliability_profile_Vol2}
\end{figure}

% --- Figure S6: Reliability profile Vol4 ---
\begin{figure}[H]
  \centering
  \includegraphics[width=\linewidth]{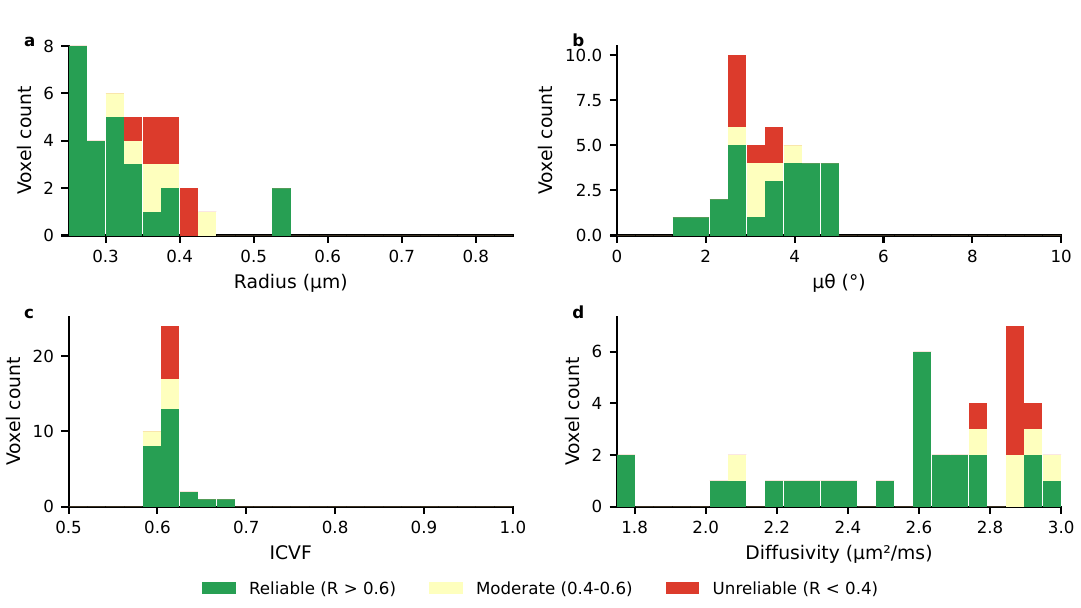}
  \caption{Reliability-layered parameter histograms for Rat~4 ($n = 38$ corpus callosum voxels), same format as Figure~\ref{fig:reliability_profile_Vol1}. Rat~4 is the least reliable animal (74\% of voxels with $R > 0.5$), with a lower median $S_{\text{match}}$ than the other animals.}
  \label{fig:supp_reliability_profile_Vol4}
\end{figure}

% --- Figure S7: Spatial maps Vol2 ---
\begin{figure}[H]
  \centering
  \includegraphics[width=\linewidth]{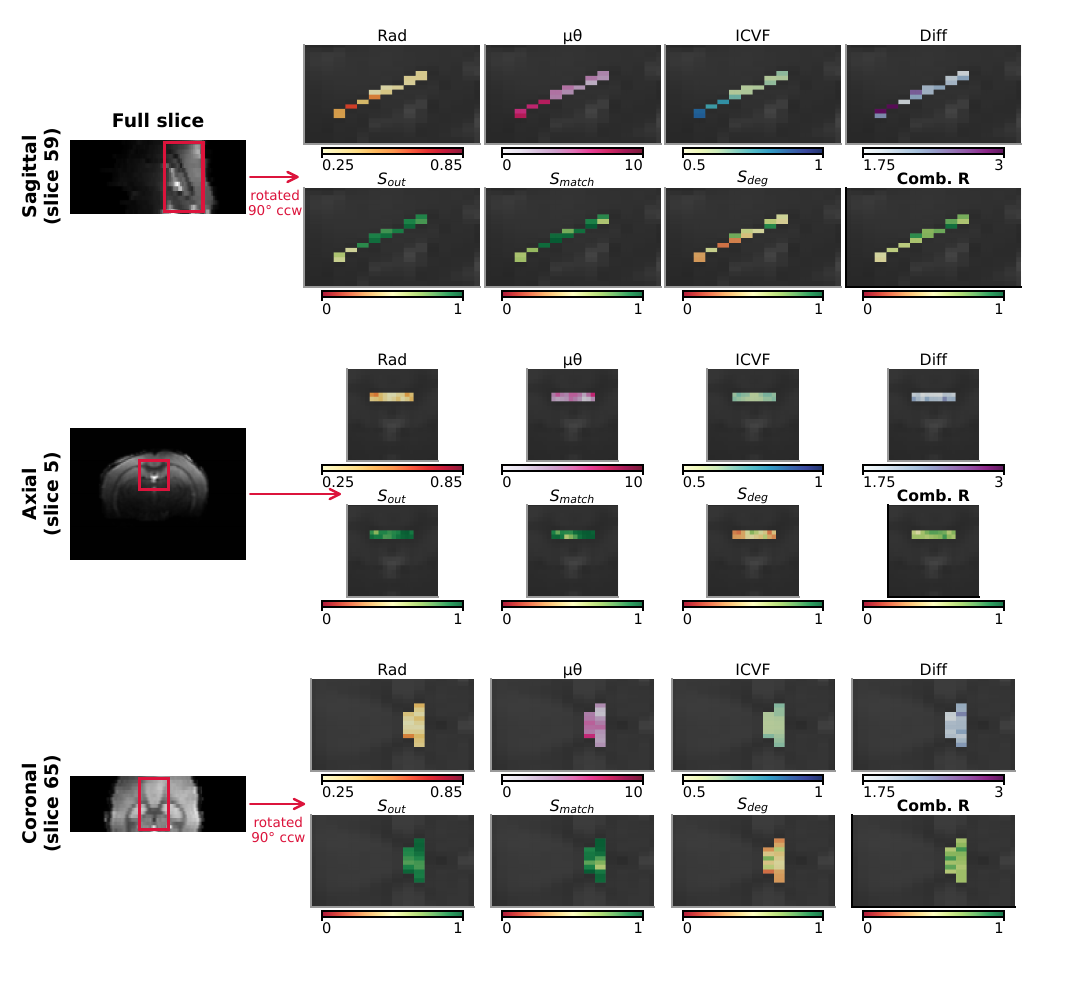}
  \caption{Spatial parameter maps for Rat~2 ($n = 89$ voxels), same format as Figure~\ref{fig:spatial_maps_Vol1}.}
  \label{fig:supp_spatial_maps_Vol2}
\end{figure}

% --- Figure S8: Spatial maps Vol4 ---
\begin{figure}[H]
  \centering
  \includegraphics[width=\linewidth]{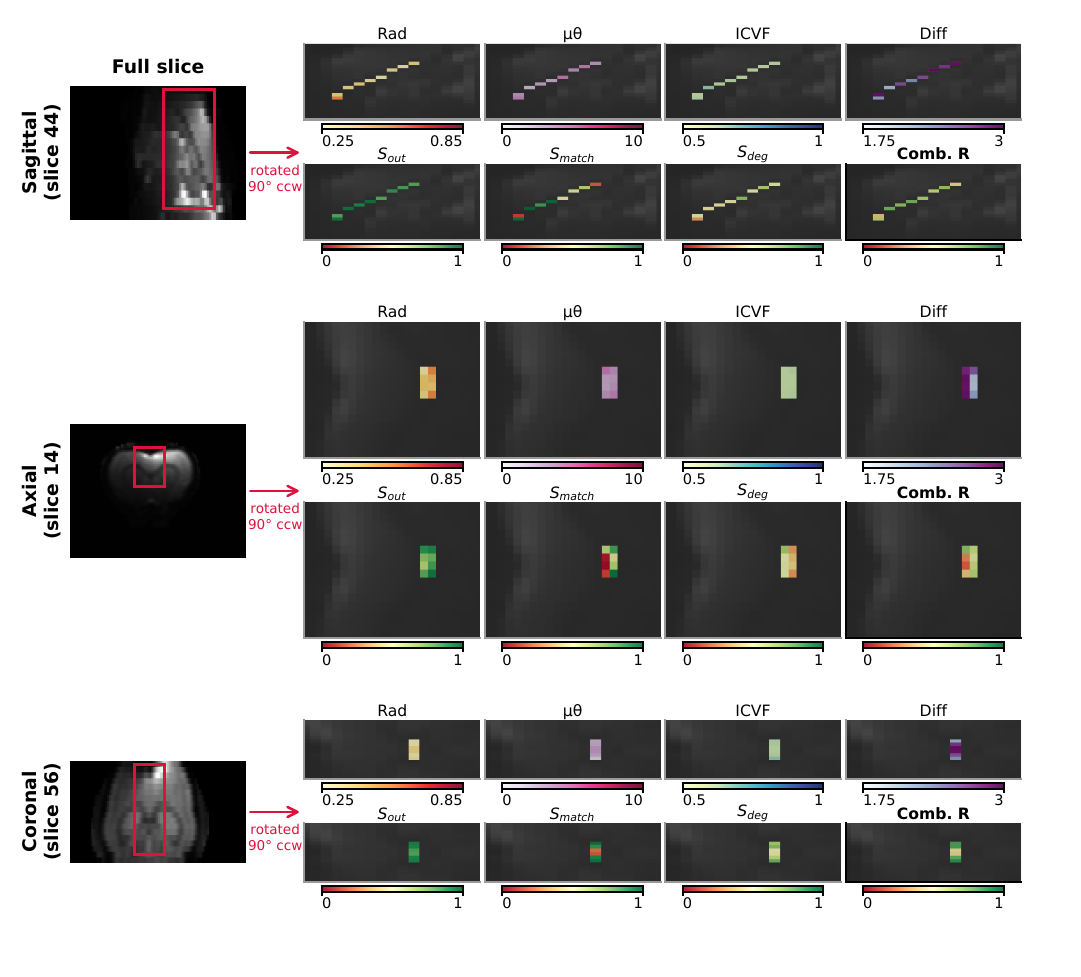}
  \caption{Spatial parameter maps for Rat~4 ($n = 38$ voxels), same format as Figure~\ref{fig:spatial_maps_Vol1}.}
  \label{fig:supp_spatial_maps_Vol4}
\end{figure}

% --- Figure S9: Complement maps Vol2 ---
\begin{figure}[H]
  \centering
  \includegraphics[width=\linewidth]{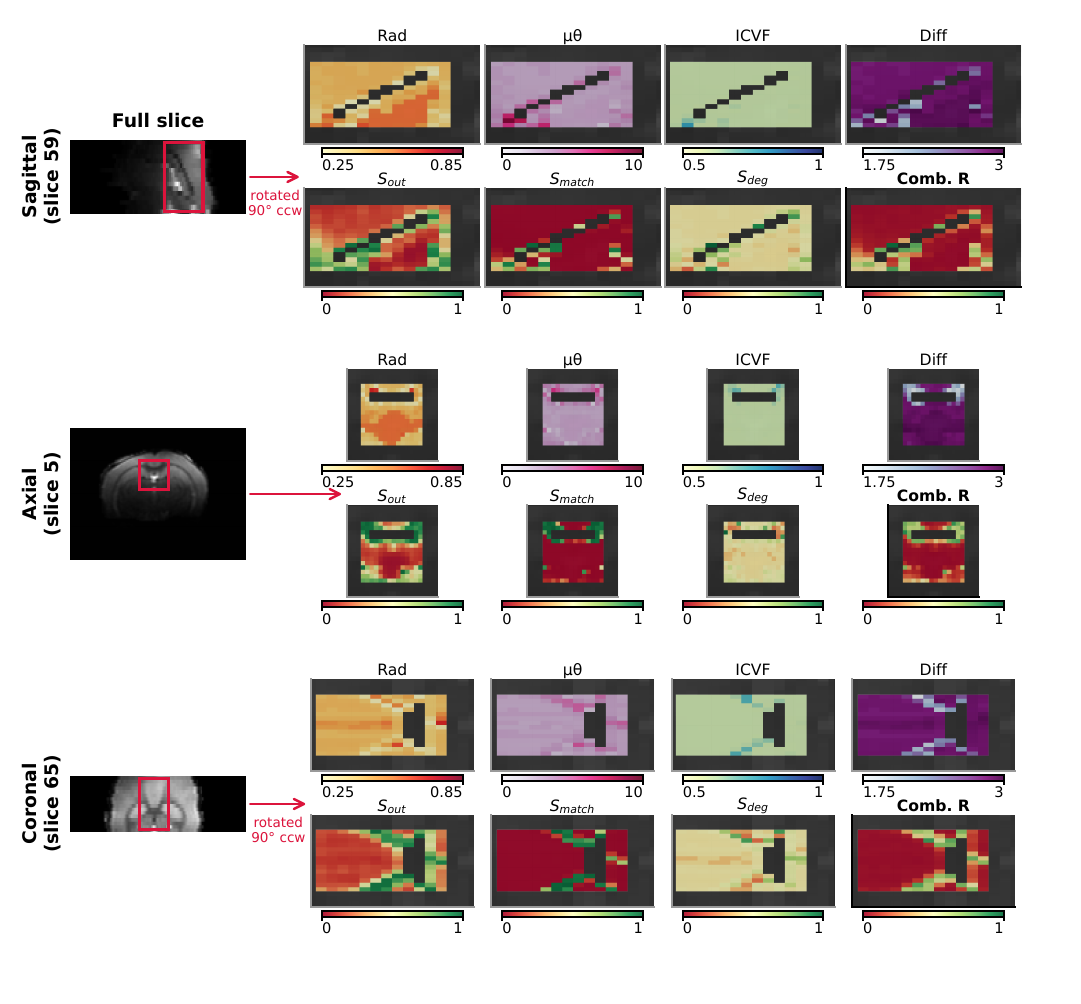}
  \caption{Complement spatial parameter maps for Rat~2, same format as Figure~\ref{fig:complement_maps_Vol1}.}
  \label{fig:supp_complement_maps_Vol2}
\end{figure}

% --- Figure S10: Complement maps Vol3 ---
\begin{figure}[H]
  \centering
  \includegraphics[width=\linewidth]{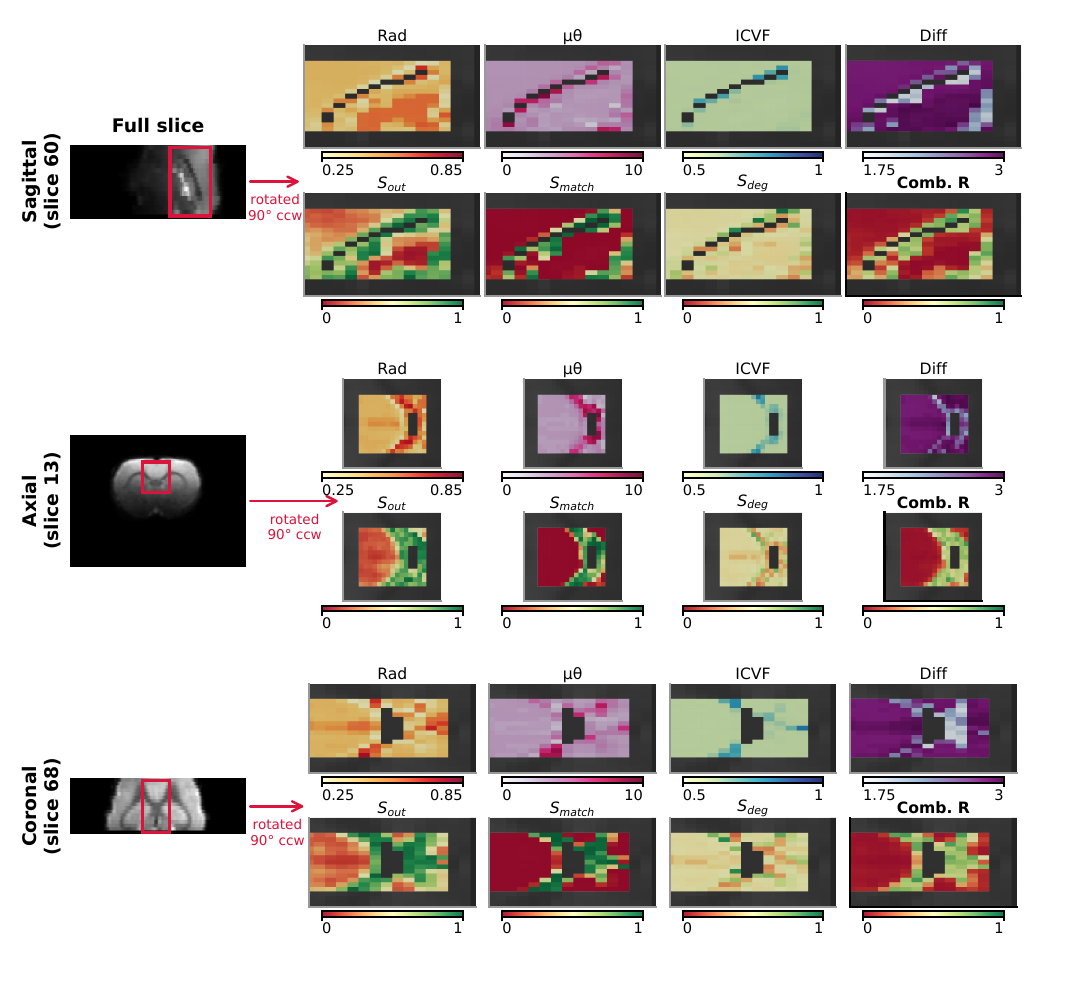}
  \caption{Complement spatial parameter maps for Rat~3, same format as Figure~\ref{fig:complement_maps_Vol1}.}
  \label{fig:supp_complement_maps_Vol3}
\end{figure}

% --- Figure S11: Complement maps Vol4 ---
\begin{figure}[H]
  \centering
  \includegraphics[width=\linewidth]{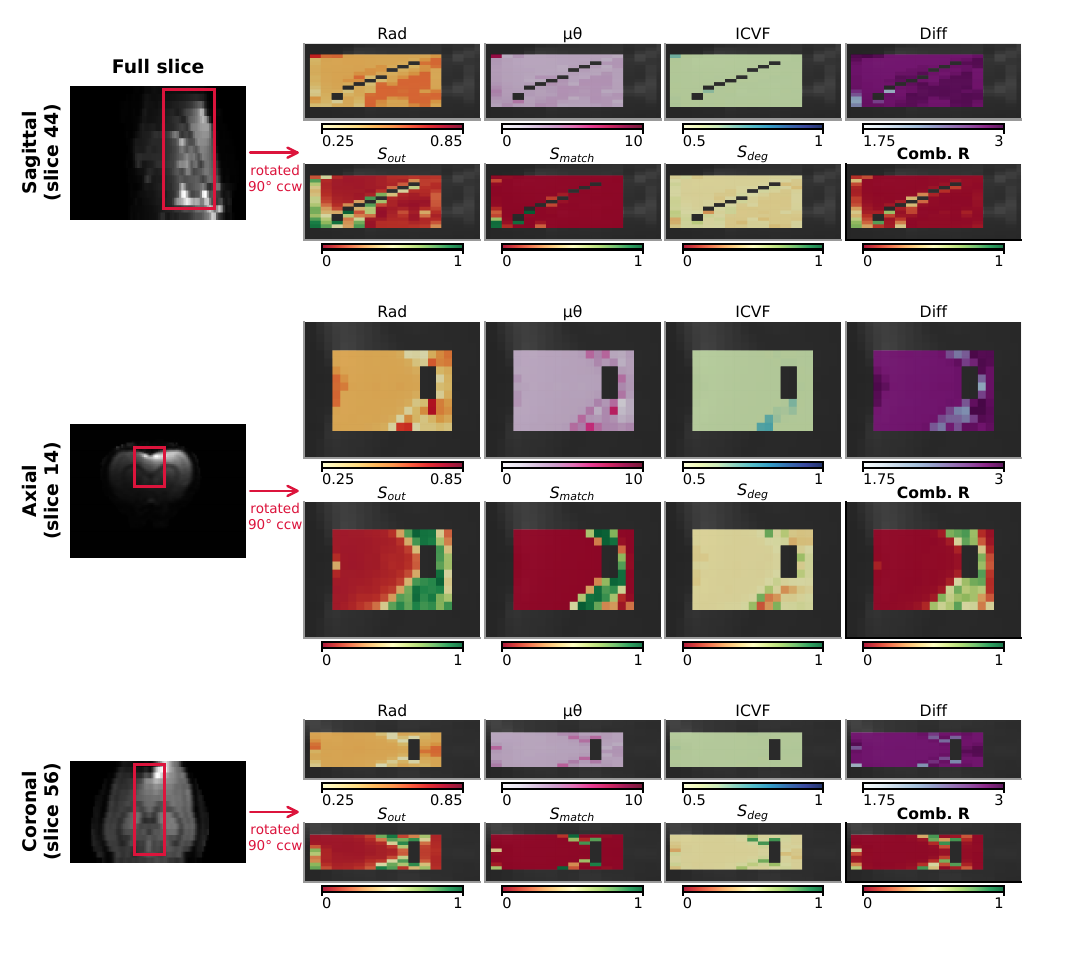}
  \caption{Complement spatial parameter maps for Rat~4, same format as Figure~\ref{fig:complement_maps_Vol1}.}
  \label{fig:supp_complement_maps_Vol4}
\end{figure}

% ---- Supplementary Table (per-voxel; superseded by the reliability-layered histograms) ----

Per-voxel estimates and reliability scores are summarised by the reliability-layered histograms (Figures~\ref{fig:reliability_profile_Vol1}, \ref{fig:reliability_profile_Vol3}, \ref{fig:supp_reliability_profile_Vol2}, and~\ref{fig:supp_reliability_profile_Vol4}) and by the per-animal medians in Table~\ref{tab:per_volume}.

\renewcommand{\arraystretch}{1.15}

% end superseded per-voxel table

% --- Bibliography ---
\bibliography{references}

% ============================================================================
% Back matter
% ============================================================================
\section*{Funding}

This work was supported by the Swiss National Science Foundation under grant 205320\_204097. Open access was funded by the \'Ecole Polytechnique F\'ed\'erale de Lausanne.

\section*{Acknowledgements}

We acknowledge access to the facilities and expertise of the CIBM Center for Biomedical Imaging, a Swiss research centre of excellence founded and supported by Lausanne University Hospital (CHUV), University of Lausanne (UNIL), \'Ecole Polytechnique F\'ed\'erale de Lausanne (EPFL), University of Geneva (UNIGE), and Geneva University Hospitals (HUG).

\section*{Author contributions statement}

J.L.V.H.\ co-designed the study, generated substrates and simulations, implemented the estimation and reliability pipeline, performed the analyses, and drafted the manuscript.

J.R.-P.\ co-designed the study, contributed to the methodological development, provided scientific discussion throughout the project, supervised the work, drafted and revised the manuscript.

I.J.\ shared the in vivo NEXI dataset, contributed to the interpretation of results, and revised the manuscript.

J.-P.T.\ supervised the project, contributed to the interpretation of results, and revised the manuscript.

All authors reviewed and approved the final manuscript.

\section*{Data availability}

The in vivo DW-MRI data used in this study were acquired as part of the NEXI study\cite{Jelescu2022} and are available from the corresponding author upon reasonable request. The CACTUS framework is publicly available at \url{https://github.com/Juanitovh/CACTUS}. The MC-DC simulator is publicly available at \url{https://github.com/jonhrafe/MCDC_Simulator_public}. The signal dictionary will be made publicly available on publication in \href{https://drive.google.com/drive/folders/1UwRqJqyZzLSxhvJ-K10URCYjqwVSKEhJ?usp=sharing}{this shared Google Drive folder}.

\section*{Competing interests}

The authors declare no competing interests.

\end{document}